\documentclass[
preprint,5p,twocolumn,final,number,
sort&compress]{elsarticle}

\usepackage{amsmath,amssymb,amsfonts}
\usepackage{longtable}
\usepackage{array,dcolumn}
\usepackage{graphicx}
\usepackage{dcolumn}
\usepackage{bm}
\usepackage{comment}
\usepackage[dvipsnames]{xcolor}
\usepackage{lineno,hyperref}
\usepackage{textcomp}
\usepackage[capitalize]{cleveref}
\usepackage{natbib}
\usepackage{geometry}
\usepackage{txfonts}
\usepackage{xcolor}

\hypersetup{pdfauthor={Name}}

\journal{Nuclear Instruments and Methods in Physics Research Section A}

\biboptions{sort&compress}

\begin{document}

\begin{frontmatter}

\title{G4CMP: Condensed Matter Physics Simulation Using the {\sc Geant4} Toolkit}

\author[1]{M.\,H.~Kelsey\corref{cor1}}
\author[2]{R.~Agnese}
\author[1]{Y.\,F.~Alam}
\author[1]{I.~Ataee~Langroudy}
\author[1]{E.~Azadbakht}
\author[3]{D.~Brandt}
\author[4]{R.~Bunker\corref{cor1}}
\author[5]{B.~Cabrera}
\author[6]{Y.-Y.~Chang}
\author[2]{H.~Coombes}
\author[7]{R.\,M.~Cormier}
\author[7]{M.\,D.~Diamond}
\author[4]{E.\,R.~Edwards}
\author[8]{E.~Figueroa-Feliciano}
\author[9]{J.~Gao}
\author[10]{P.\,M.~Harrington}
\author[7]{Z.~Hong}
\author[7]{M.~Hui}
\author[3]{N.\,A.~Kurinsky}
\author[1]{R.\,E.~Lawrence}
\author[4]{B.~Loer}
\author[12]{M.\,G.~Masten}
\author[13]{E.~Michaud}
\author[14,15]{E.~Michielin}
\author[4]{J.~Miller}
\author[8]{V.~Novati}
\author[4]{N.\,S.~Oblath}
\author[4]{J.\,L.~Orrell}
\author[1]{W.\,L.~Perry}
\author[5]{P.~Redl}
\author[7]{T.~Reynolds}
\author[2]{T.~Saab}
\author[16,17]{B.~Sadoulet}
\author[10,18]{K.~Serniak}
\author[5]{J.~Singh}
\author[2]{Z.~Speaks}
\author[19]{C.~Stanford}
\author[9]{J.\,R.~Stevens}
\author[4,20]{J.~Strube}
\author[1]{D.~Toback}
\author[9,21]{J.\,N.~Ullom}
\author[4]{B.\,A.~VanDevender}
\author[21]{M.\,R.~Vissers}
\author[22]{M.\,J.~Wilson}
\author[23]{J.\,S.~Wilson}
\author[7]{B.~Zatschler}
\author[7]{S.~Zatschler}

\address[1]{Department of Physics and Astronomy, and the Mitchell Institute for Fundamental Physics and Astronomy, Texas A\&M University, College Station, TX 77843, USA}
\address[2]{Department of Physics, University of Florida, Gainesville, FL 32611, USA}
\address[3]{SLAC National Accelerator Laboratory/Kavli Institute for Particle Astrophysics and Cosmology, Menlo Park, CA 94025, USA}
\address[4]{Pacific Northwest National Laboratory, Richland, WA 99352, 
USA}
\address[5]{Department of Physics, Stanford University, Stanford, CA 94305, USA}
\address[6]{Division of Physics, Mathematics, \& Astronomy, California Institute of Technology,Pasadena, CA 91125, USA}
\address[7]{Department of Physics, University of Toronto, Toronto, ON M5S 1A7, Canada}
\address[8]{Department of Physics \& Astronomy, Northwestern University, Evanston, IL 60208, USA}
\address[9]{University of Colorado, Department of Physics, Boulder, CO 80309, USA}
\address[10]{Research Laboratory of Electronics, Massachusetts Institute of Technology, Cambridge, MA 02139, USA}
\address[12]{Department of Physics, Santa Clara University, Santa Clara, CA 95053, USA}
\address[13]{D\'{e}partement de Physique, Universit\'{e} de Montr\'{e}al, Montr\'{e}al, QC H3C 3J7, Canada}
\address[14]{Department of Physics \& Astronomy, University of British Columbia, Vancouver, BC V6T 1Z1, Canada}
\address[15]{TRIUMF, Vancouver, BC V6T 2A3, Canada}
\address[16]{Department of Physics, University of California, Berkeley, CA 94720, USA}
\address[17]{Lawrence Berkeley National Laboratory, Berkeley, CA 94720, USA}
\address[18]{Lincoln Laboratory, Massachusetts Institute of Technology, Lexington, MA 02421, USA}
\address[19]{Fermi National Accelerator Laboratory, Batavia, IL 60510, USA}
\address[20]{Institute for Fundamental Science, University of Oregon, Eugene, OR 97403, USA}
\address[21]{National Institute of Standards and Technology, Quantum Electromagnetics Division, Boulder, CO 80305, USA}
\address[22]{Institute for Astroparticle Physics (IAP), Karlsruhe Institute of Technology (KIT), 76344, Germany}
\address[23]{Department of Physics, Baylor University, Waco, TX 76798, USA}

\cortext[cor1]{Corresponding: kelsey@slac.stanford.edu, 
raymond.bunker@pnnl.gov}

\begin{abstract}
G4CMP simulates phonon and charge transport in cryogenic semiconductor crystals using the {\sc Geant4} toolkit. The transport code is capable of simulating the propagation of acoustic phonons as well as electron and hole charge carriers. Processes for anisotropic phonon propagation, oblique charge-carrier propagation, and phonon emission by accelerated charge carriers are included. The simulation reproduces theoretical predictions and experimental observations such as phonon caustics, heat-pulse propagation times, and mean charge-carrier drift velocities. In addition to presenting the physics and features supported by G4CMP, this report outlines example applications from the dark matter and quantum information science communities. These communities are applying G4CMP to model and design devices for which the energy transported by phonons and charge carriers is germane to the performance of superconducting instruments and circuits placed on silicon and germanium substrates. The G4CMP package is available to download from GitHub: \href{https://github.com/kelseymh/G4CMP}{github.com/kelseymh/G4CMP}.
\end{abstract}

\end{frontmatter}


\section{\label{sec:intro}Introduction}
The {\sc Geant4} Condensed Matter Physics (G4CMP) package was introduced and released approximately 10 years ago~\cite{Brandt,arXiv:1403.4984} as a publicly available addition to the {\sc Geant4} toolkit. {\sc Geant4} finds application in ``high energy, nuclear and accelerator physics, as well as studies in medical and space science''~\cite{Geant-A,Geant-B,Geant-C} as a tool for simulating the passage of particles through matter. Both {\sc Geant4} and G4CMP are written in the {C/C++} programming language. G4CMP comprises a set of generalized physics processes related to the production and transport at cryogenic temperatures ($T\ll1$\,K) of nonequilibirium solid-state excitations:  phonons, electron-hole pairs, and Bogoliubov quasiparticles.\footnote{Hereafter in this paper, when we use the term ``quasiparticles'' we are referring specifically to Bogoliubov quasiparticles (see, \textit{e.g.}, Refs.~\cite{PhysRevB.41.11693,Ronen2016,10.21468/SciPostPhysLectNotes.31} and references therein), which result from breaking of Cooper pairs in superconductors.} The package was originally devised and deployed to model the response of a specific class of devices for direct detection of dark matter in the Cryogenic Dark Matter Search (CDMS) experiment~\cite{CDMS-A,SuperCDMS:2014kpk}. Today, 
G4CMP is being adopted for analyses of energy injection and transport in a broad class of cryogenic devices, including next-generation dark-matter detectors~\cite{SuperCDMS:2016wui,Ren:2020gaq,SuperCDMS:2020aus,SuperCDMS:2020ymb} 
as well as superconducting quantum sensing and quantum computing devices (see, \textit{e.g.}, Refs.~\cite{Wilen:2020lgg,Martinez:2018ezx}). These quantum-based devices share similar  
substrate material and superconducting circuitry with their dark-matter detector cousins.

The {\sc Geant4} toolkit is an event-based, Monte Carlo-driven simulation engine. The toolkit provides modeling of arbitrary geometries composed of volumes with defined properties and also includes the ability to simulate radiation sources and track particle interactions throughout a geometry. To this framework, the G4CMP package adds phonon and charge modeling in solid state volumes, thus providing a bridge between {\sc Geant4}'s particle physics and solid-state detector response. 
In simplified terms, G4CMP provides a method for modeling charge and athermal phonon production and transport in a material (\textit{e.g.}, semiconductor lattice) in response to the passage of particles through the material. This modeling and simulation apparatus is thus developed around a concept of material and device response to specified incident radiation.  
Notably, equilibrium (\textit{i.e.}, ``thermal'') processes are not explicitly modeled via the Monte Carlo transport method; however, they could, in principle, be approximated with additional G4CMP physics processes within the {\sc Geant4} framework. 

Our goal with this article is to update the description of the G4CMP package, which has evolved from the Monte Carlo techniques originally described in Refs.~\cite{leman,Brandt,arXiv:1403.4984}.\footnote{This article presents the status of G4CMP for package version 8.2 under {\sc Geant4} version 10.07.p03, which are each available online for download~\cite{G4CMP_code_repo,GEANT4_code_repo}.  Documentation on GitHub at Ref.~\cite{G4CMP_code_repo} indicates how the community may contribute to the package if desired.}
This includes covering details of the physics processes supported by the package (Sec.~\ref{sec:processes}). For practical use, some of the important features and functionality of G4CMP are outlined (Sec.~\ref{sec:features}). 
Finally, with the recent broadening of the use of the package, concise examples are provided to demonstrate the breadth of application (Sec.~\ref{sec:examples}). Section~\ref{sec:summary} provides a summary and a brief discussion of desired features that may be pursued in future development of the package.


 \section{\label{sec:processes}Physics processes modeled in G4CMP}
The physics processes modeled in G4CMP enable {\sc Geant4} tracking of production and transport of nonequilibrium phonons and charges in solid-state materials. As of this writing, the official G4CMP release includes processes for such tracking in germanium and silicon semiconductor lattices, with the option to include charge and phonon sensors (or ``electrodes'') on the surfaces. 

These processes capture a class of solid-state devices in which the surfaces of single-crystal semiconductor substrates are instrumented with superconducting circuits. This includes devices explicitly designed for sensitivity to energy absorption within the substrate or superconducting circuits, such as dark-matter detectors~\cite{SuperCDMS:2014kpk}, photon sensors using microwave kinetic-inductance detectors (MKIDs)~\cite{WOS:000645772700001}, and superconducting-nanowire single-photon detectors (SNSPDs)~\cite{WOS:000706890200035,WOS:000304048100003}. Also included are devices whose performance benefits from insensitivity to environmental disturbances, such as superconducting qubits affected by ionizing radiation~\cite{Vepsalainen:2020trd,Wilen:2020lgg,McEwen:2021wdg}.
In this section, we review the physics processes modeled in G4CMP that enable tracking of nonequilibrium excitations in this broad class of devices. The majority of the section describes phonon and charge production and transport in semiconductor substrates.
We also describe the superconductor processes that G4CMP currently includes to enable simulation of quasiparticle-sensitive sensors.

\begin{figure*}[t!]
	\centering
	\includegraphics[trim={1.8cm 2.3cm 4.2cm 0.3cm},clip,width=0.8\textwidth]{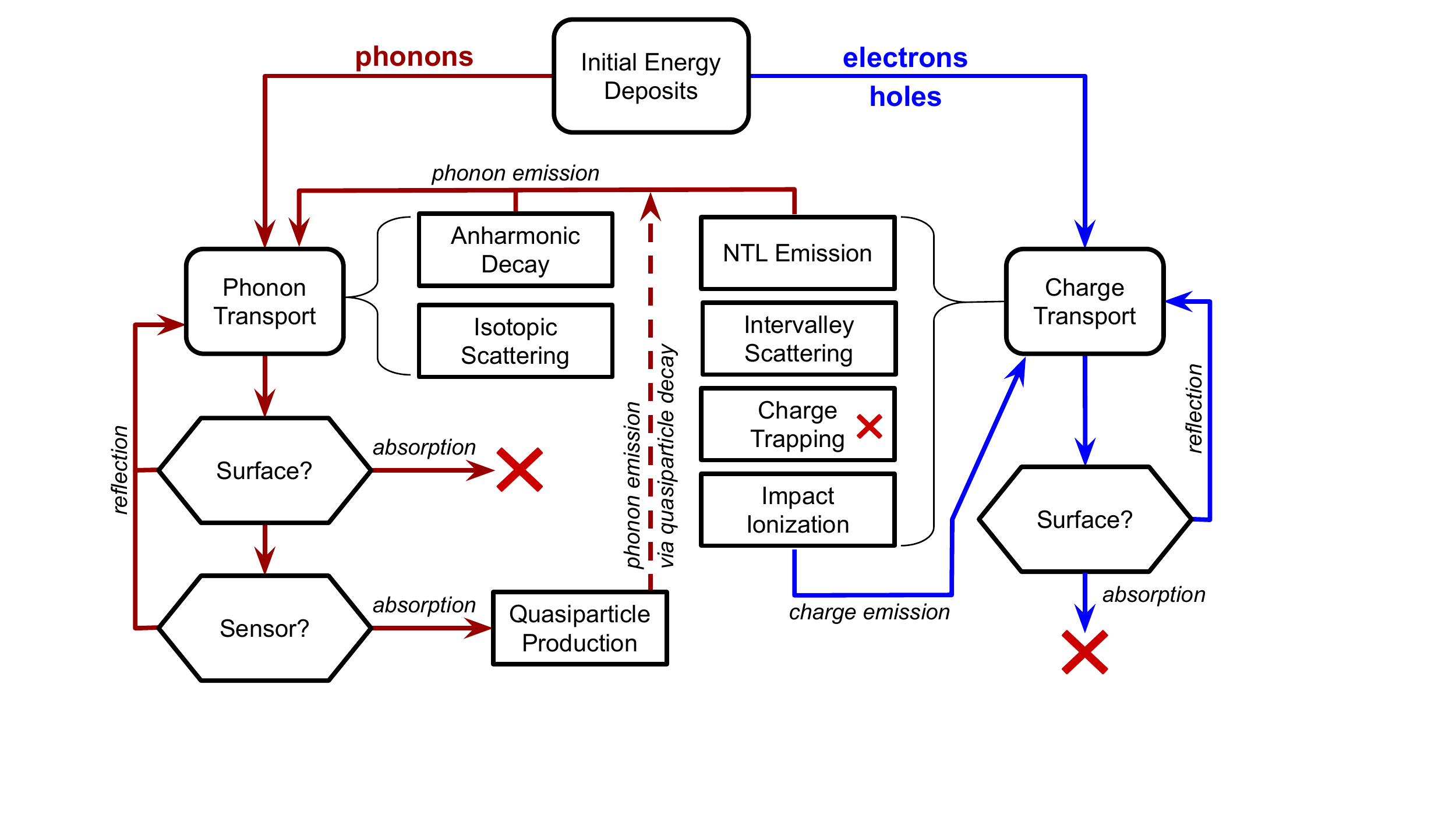}
    \caption{\label{fig:flowchart}Overall event flow of G4CMP processing in {\sc Geant4}.
    As described in Sec.~\ref{sec:Production}, energy deposits (top) from {\sc Geant4} are converted into primary phonons (left path) and/or electron-hole pairs (right path). 
    These particles are transported through the substrate material with appropriate interactions, as listed in each column.
    The processes of phonon anharmonic decay (Sec.~\ref{sec:PhononProcesses}), Neganov-Trofimov-Luke (NTL) emission (Sec.~\ref{sec:NTL}), impact ionization (Sec.~\ref{sec:ct-ii}), and quasiparticle production (Sec.~\ref{sec:Quasiparticles}) can create additional particles, which are added to the processing loop as indicated by arrows.  The dashed arrow indicates that the quasiparticle-production process may or may not create any phonons at a given step. 
    When a surface is encountered (Sec.~\ref{sec:surfaces}), phonons (lower left) may continue to be transported after reflection or be absorbed on a sensor; charges may reflect or be absorbed.}
\end{figure*}

Figure~\ref{fig:flowchart} shows the overall flow of the particles and processes implemented in G4CMP, in the context of the {\sc Geant4} event loop, with phonons and/or electron-hole (e$^-$/h$^+$) pairs being created from energy deposits recorded as a result of other particle interactions.  Phonons and e$^-$/h$^+$ pairs may also be created directly as primary particles,  which is a useful option in some applications (see, \textit{e.g.}, Sec.\ \ref{sec:qischip}).  
In either case, the phonons and e$^-$/h$^+$ pairs are transported through the user's geometry and interact via the processes described in this section. At surfaces, phonons may be reflected and continue their transport, or they may be absorbed at a defined sensor (see Sec.~\ref{sec:surfaces}); charges are terminated at surfaces by default.

\subsection{\label{sec:Production}Phonon and charge production}

{\sc Geant4} typically records either the energy deposited by a particle or the secondary tracks produced by an interaction. 
However, to support different simulation strategies, G4CMP does not enforce an exclusive categorization of the kind of recorded information.  This flexibility is beneficial for implementation of the G4CMP physics processes. When a {\sc Geant4} energy deposit is recorded, G4CMP uses the type of interaction -- electromagnetic or screened nuclear recoil -- and the amount of energy deposited to compute how many e$^-$/h$^+$ pairs and phonons would be created. These charges and phonons are then added as secondary particles which are subsequently tracked. Nevertheless, the value of the originally deposited energy in the interaction is preserved for later analysis.

In semiconductor substrates, an e$^-$/h$^+$ pair is produced when some of the deposited energy promotes an electron into the conduction band, leaving a hole behind.
Creation of an e$^-$/h$^+$ pair requires a minimum energy corresponding to the material's bandgap. 
On average, however, additional energy -- typically three times the bandgap~\cite{Fang:TNS2018} -- is needed such that the electron and hole have sufficient kinetic energy to prevent them from immediately recombining.  
This additional $\mathcal{O}$(eV) energy is large enough to scatter phonons and produce prompt phonon emission, which we model using the charge-phonon scattering process described in Sec.~\ref{sec:NTL}. 
The number of e$^-$/h$^+$ pairs produced in any given energy deposit has a Poisson-like distribution, with a Fano factor~\cite{PhysRev.72.26} specific to the substrate material~\cite{WOS:A1997YF19500017,RODRIGUES2021165511}. The ionization model currently implemented in G4CMP has a Fano factor and yield that are independent of energy. Future work will incorporate the energy-dependent model of Ref.~\cite{Ramanathan_2020} when data are available to calibrate the energy dependence in the Fano factor. 

For interactions from photons or high-energy electrons, the full energy deposit is converted directly into e$^-$/h$^+$ pairs.  
For interactions involving ions (including $\alpha$-particles) or recoiling nuclei, a portion of the deposited energy is transferred directly into production of lattice vibrations (prompt phonons),
while the remainder goes into creation of e$^-$/h$^+$ pairs. The ratio of energies -- production of e$^-$/h$^+$ pairs versus creation of prompt phonons -- depends on the total deposited energy.  In G4CMP, this ratio is determined by an ionization yield model; two  models from Refs.~\cite{Lindhard,LewinSmith,Robinson} are currently implemented (see Sec.\ \ref{sec:physicsconfig}).

\subsection{\label{sec:PhononTransport}Phonon transport}
Phonon transport was the first component of the G4CMP framework to be developed
\cite{Brandt}. The phonon-transport code described here is intended for
temperatures $T\ll1$\,K; thus, transport and scattering of thermally excited background phonons is ignored. Currently, only acoustic phonons are simulated in G4CMP. Optical phonons are not yet supported. In millikelvin devices, 
optical phonons immediately down-convert to lower-energy (acoustic) phonon modes. This exclusion of optical phonons means G4CMP does not support direct interaction with electromagnetic (photon) phenomena (\textit{e.g.}, photon-phonon scattering).

\subsubsection{\label{sec:Focusing}Anisotropic transport and phonon focusing}
As phonons are quantized vibrations of the crystal lattice, the propagation of phonons is governed by a three-dimensional wave equation \cite{Wolfe}:
\begin{equation}
\label{eq:3DWave}
\rho \omega^2 e_i = C_{ijml} k_j k_m e_l\ ,
\end{equation}
where $\rho$ is the crystal mass density, $\omega$ is the angular phonon frequency, $\vec{e}$ is the polarization vector, $\vec{k}$ is a wavevector, $C_{ijml}$ is the elasticity tensor, and there is an implicit sum over indices $j$, $m$ and $l$.

\begin{figure*}[t!]
	\centering
	\includegraphics[height=1.4in]{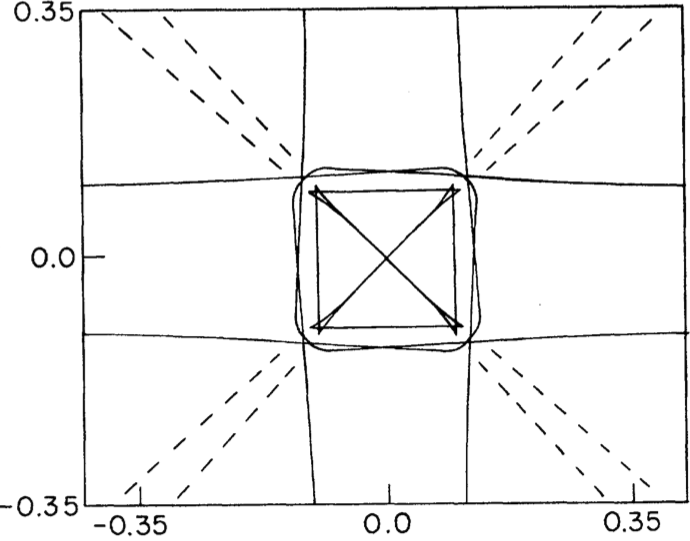}\qquad
	\raisebox{-.15in}{\includegraphics[height=1.75in]{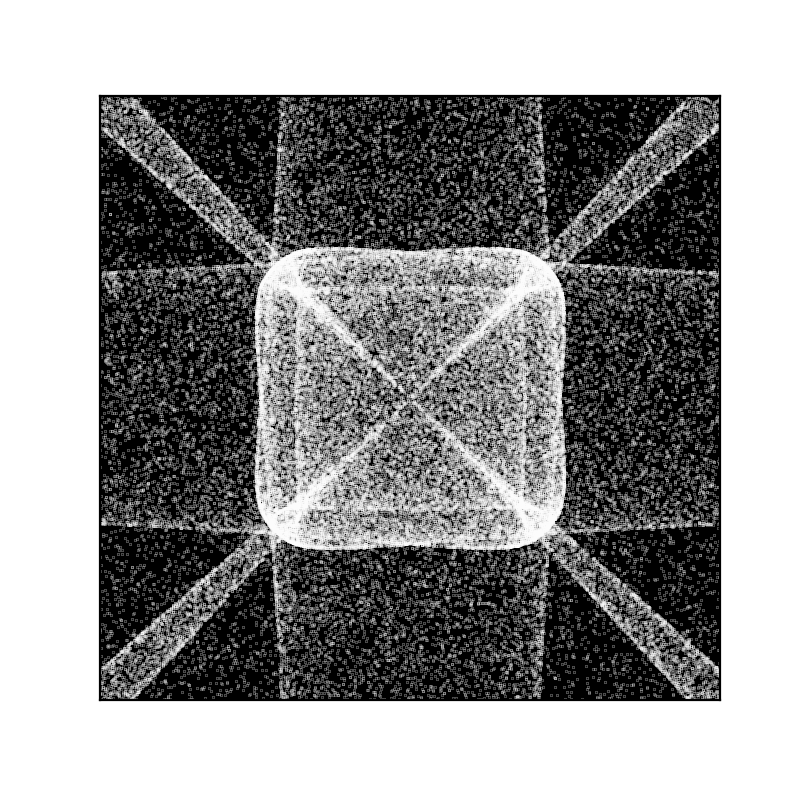}\qquad
	\includegraphics[height=1.75in]{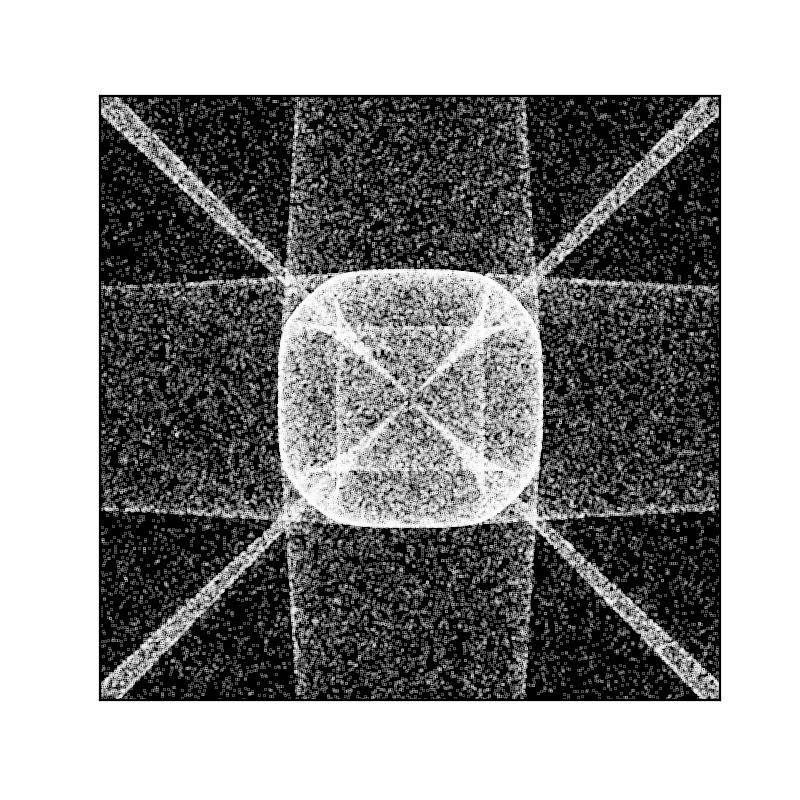}}
	\caption{
	Comparison of phonon caustics predicted for a point source in a 1\,cm thick Ge $\langle$100$\rangle$ crystal with corresponding results from G4CMP, showing positions of transverse phonon modes on the face opposite the point source.
	\textbf{Left:} Outline of phonon caustics in Ge $\langle$100$\rangle$ as predicted by Nothrop and Wolfe \cite{Nothrop}. 
	\textbf{Middle:} Caustics pattern as simulated using G4CMP for phonon transport in Ge $\langle$100$\rangle$, in good agreement with the theoretical prediction to the left. 
	\textbf{Right:} Caustics pattern as simulated using G4CMP for phonon transport in a 1\,cm thick Si $\langle$100$\rangle$ crystal (see also Ref.~\cite{Martinez:2018ezx}).}
	\label{fig:caustics}
\end{figure*}

For any given wavevector $\vec{k}$, Eq.~\ref{eq:3DWave} has three
eigenvalues $\omega$ and three eigenvectors $\vec{e}$. 
These correspond to the three different phonon polarization states (or acoustic modes): longitudinal, fast transverse, and slow transverse. 
The actual direction and
velocity of propagation of phonons is given by the group velocity 
$\vec{v}_g$, calculated by interpreting $\omega$ in Eq.\ \ref{eq:3DWave} as a function of $\vec{k}$:
\begin{equation}
\label{eq:GroupV}
\vec{v}_g=\nabla_k \omega (\vec{k}).
\end{equation}
The group velocity $\vec{v}_g$ is not parallel to the phonon momentum
$\hbar\vec{k}$ because of anisotropy in $C_{ijml}$. Instead, phonons are focused into propagation directions that
correspond to the highest density of eigenvectors. This focusing
gives rise to caustics in the spatial distribution of phonons in the substrate as they are transported away from a
point-like source that is isotropic in $\vec{k}$-space. The resulting caustics can be observed using micro-calorimeters \cite{Nothrop,Jakata}. Figure~\ref{fig:caustics} shows that the caustics simulated with the G4CMP phonon-transport code are in good agreement with expectations from the work of Nothrop and Wolfe \cite{Nothrop}. 

The wave equation is not solved in real time in the G4CMP transport code. Instead, a look-up table for Eq.~\ref{eq:GroupV} is generated that maps
$\vec{k}$ onto $\vec{v}_g$, and bilinear interpolation is used to generate a continuous mapping function. Phonon focusing and methods for solving the three-dimensional wave equations are treated in Ref.~\cite{Wolfe}.

\subsubsection{\label{sec:PhononProcesses}Phonon scattering processes}
In addition to the anisotropic transport described by Eq.~\ref{eq:3DWave}, G4CMP also includes modeling of two other processes relevant to
acoustic-phonon transport in cryogenic crystals: \textit{anharmonic down-conversion} and \textit{isotopic scattering} \cite{Tamura1,Tamura2,Tamura3}. The rates of these processes are given by \cite{Brandt, Tamura2}:
\begin{equation}
\label{eq:phononRates}
\begin{split}
\Gamma_{\mathrm{anh}} &= A\nu^5,\\ 
\Gamma_{\mathrm{scatter}} &= B\nu^4,
\end{split}
\end{equation}
where $\Gamma_{\mathrm{anh}}$ is the number of anharmonic down-conversion events per unit time,   
$\Gamma_{\mathrm{scatter}}$ is the number of isotopic scattering events per unit time,
$\nu$ is the phonon frequency, and $A$ and $B$ are constants of proportionality
related to the crystal's elasticity tensor. 
In G4CMP, the current default values of $A$ and $B$ for Ge and Si are taken from Ref.~\cite{Tamura3}.  
Additional parameters used to simulate phonon transport are listed in \ref{sec:matconfig} (see also Ref.~\cite{leman}).

Isotopic scattering occurs when a phonon interacts with an isotopic substitution
site in the lattice. It is effectively an elastic scattering event during which the
phonon momentum vector is randomized and the polarization state can change freely among the acoustic modes. 
The partition among the polarization states is determined by the relative density of allowed states. 
This change among polarization states is often referred to as \emph{mode mixing} and results in the steady-state phonon populations in Ge (Si) of 53.5\% (53.1\%) slow transverse, 36.7\% (37.6\%) fast transverse, and 9.8\% (9.3\%) longitudinal~\cite{Tamura4}.

Anharmonic down-conversion occurs when a single phonon decays into two phonons, each with less energy than the progenitor. This process conserves energy but not momentum in the phonon system, because momentum is exchanged with the crystal lattice.  
The down-conversion rate of longitudinal (L) phonons dominates the energy evolution of the phonon system; down-conversion events from the other polarization states are negligible \cite{Tamura2} and are not implemented in G4CMP. There are two branches implemented in the down-conversion process: L~$\rightarrow$~L$^{\prime}$T and L~$\rightarrow$~TT, with energies for the outgoing longitudinal (L$^{\prime}$) and transverse (T) phonons drawn from built-in probability distributions~\cite{Tamura3}.

Equation \ref{eq:phononRates} indicates that phonon scattering rates  
strongly depend on the phonon energy $\hbar \nu$. Thus, high-energy phonons with $\nu$ on the
order of THz (or larger) start out in a diffusive regime with relatively high rates of isotopic scattering and 
anharmonic down-conversion; mean free paths are on the order of microns. After a few down-conversion events have occurred, phonon mean free paths increase to be on the
order of centimeters, which is comparable to the typical size of a cryogenic calorimeter or quantum device. This transition from diffuse to ballistic phonon transport is commonly referred to as ``quasi-diffuse'' and represents an important timescale for the initial evolution of phonon energy in the crystal. 
Simulation with G4CMP of this initial time evolution has been shown to follow real-world device performance, \textit{e.g.}, with the phonon sensor pulses in Ref.~\cite{Brandt}.  
Anharmonic down-conversion and isotopic scattering are well-understood processes and are discussed in great detail in the literature \cite{Tamura1,Tamura2,Wolfe,Tamura3}. Nevertheless, care should be taken in defining the G4CMP scattering parameters, which are generally substrate dependent.

\subsection{\label{sec:ChargeTransport}Charge transport}
In addition to phonon transport, the G4CMP framework enables simulation of charge propagation in semiconductor crystals. In this section we describe the processes implemented in G4CMP for Ge and Si: oblique propagation and intervalley scattering~\cite{IV1,IV2,IV3,IV4}, acceleration of charge carriers by an applied electric field and emission of acoustic phonons via the Neganov-Trofimov-Luke (NTL) process~\cite{NTL1,NTL2,NTL3}, and charge trapping and impact ionization~\cite{CT1,CT2}. We also include an assessment of charge-carrier drift speeds that result from simulations of these transport processes in G4CMP.

\subsubsection{\label{sec:InterValley}Oblique transport \& intervalley scattering}
\begin{figure*}[t!]
	\centering
	\includegraphics[width=0.9\textwidth]{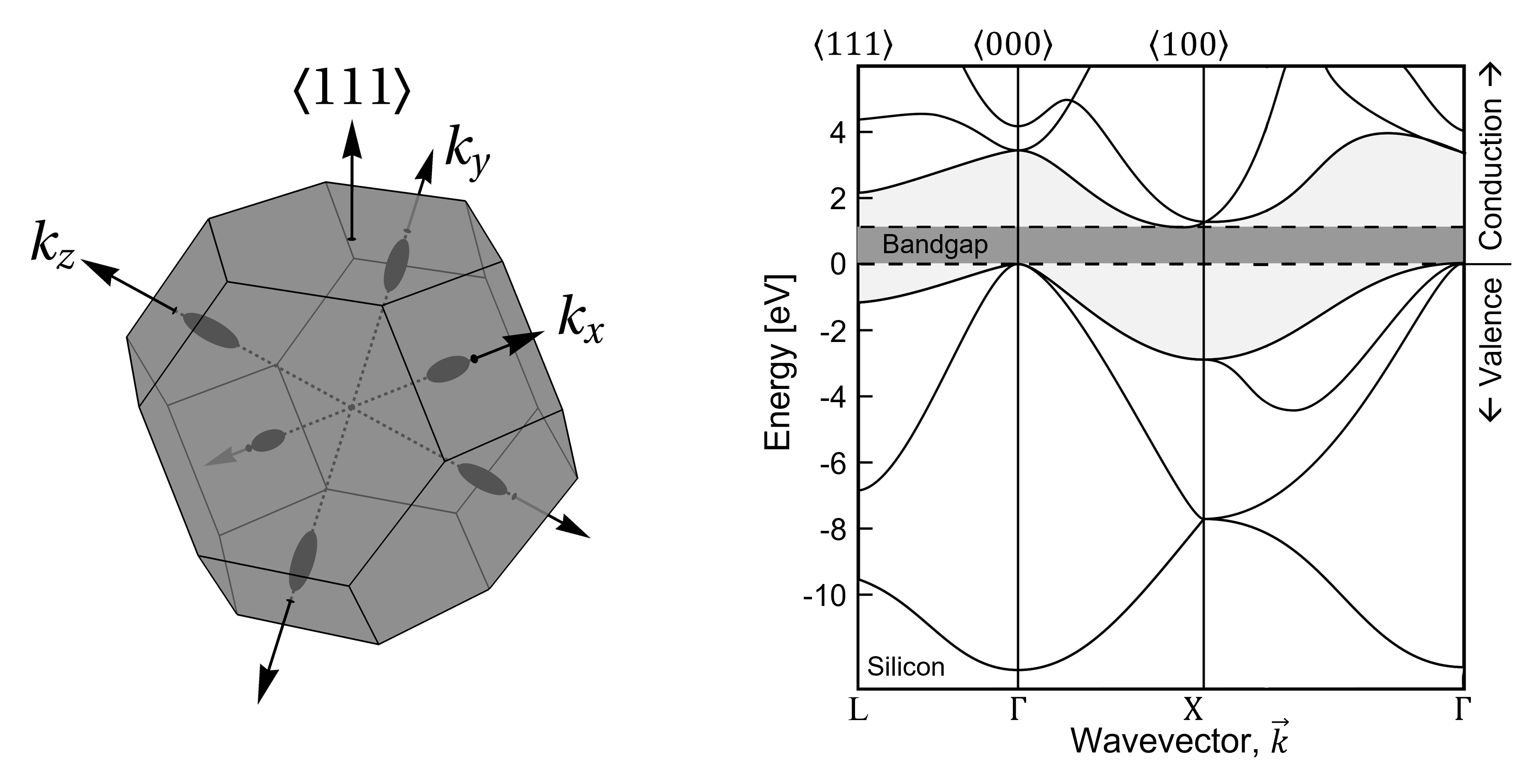}
	\caption{\textbf{Left:} The first Brillouin zone for Si, reproduced from Ref.~\cite{SiIV1}. The minimum-energy valleys of the conduction band are indicated by the ellipsoids near the edge of the Brillouin zone along the $k_x$, $k_y$, and $k_z$ directions. \textbf{Right:} Energy band structure for Si (adapted from Ref.~\cite{PhysRevB.10.5095}), highlighting the bandgap between the valence-band maximum along the $\langle$000$\rangle$ direction and the conduction-band minimum along the $\langle$100$\rangle$ direction. The latter corresponds to the minimum-energy X valleys (ellipsoids in the left panel). Similar diagrams can be found in Ref.~\cite{ashcroft} for Ge, which has its conduction-band minimum along the $\langle$111$\rangle$ direction (L valleys).}
	\label{fig:Sibands}
\end{figure*}

At the small electric fields and temperatures of interest, electron propagation in Ge and Si has the interesting feature that electrons propagate through the crystal along distinct minimum-energy valleys that are displaced from zero momentum. 
As a result, electron transport is generally oblique to the direction of the applied electric field \cite{leman,Cabrera}. 
The minimum-energy valleys  
and the energy band structure are illustrated in Fig.~\ref{fig:Sibands} for silicon, which has its conduction-band minimum along the $\langle100\rangle$ crystal axis (X valleys). In contrast, the conduction-band minimum for germanium is along the $\langle111\rangle$ direction (L valleys).

Electrons can scatter between valleys
via a process known as {\it intervalley scattering}, which occurs in one of two ways: an electron scatters off the lattice or off an impurity \cite{iv}.
The rate for both processes is dependent on the electric-field strength, with lattice scattering being the dominant factor in larger fields ($\gtrsim$5\,V/cm) and impurity scattering dominant in smaller fields ($\sim$1\,V/cm). 
G4CMP provides two parameterizations of this dependence which may be selected at runtime (see Table~\ref{tab:physics_config}, \texttt{G4CMP\_IV\_RATE\_MODEL}): a quadratic form from Ref.~\cite{GeIV},
\begin{equation}\label{eq:ivquad}
  \nu_{\rm IV} = A \left(E_0^2 + |\vec E|^2\right)^{\alpha/2}\mathrm{,}
\end{equation}
or a linear form motivated and explored in Ref.~\cite{kurinskyThesis},
\begin{equation}\label{eq:ivlinear}
  \nu_{\rm IV} = b + m|\vec E|^\alpha\mathrm{,}
\end{equation}
where $\vec E$ is the electric field, and the coefficients and exponents are specified in the ``lattice material'' table (see Table~\ref{tab:intervalley_model_config}). 

We note that the spectrum of phonons resulting from intervalley scattering at large fields may not be accurately reproduced in G4CMP.  A more complete model of the scattering process would include emission of an optical phonon and a corresponding rotation of the electron's wavevector. G4CMP does not currently model the optical phonon and thus crystal momentum is not conserved.  Instead, the electron immediately aligns itself to the new valley via the electron-phonon scattering process described in Sec.~\ref{sec:NTL}, resulting in a spectrum of acoustic phonons that is generally different from the spectrum that would result from the decay of the optical phonon. This inaccuracy could be fixed in a future update of G4CMP by modifying the intervalley scattering process to rotate the electron wavevector and create acoustic phonons based on the spectrum expected from down-conversion of the optical phonon. For small  fields, the current G4CMP implementation  
is a good representation of the effectively random valley changes that occur when electrons scatter off impurities.  

\begin{figure*}
	\centering
	\includegraphics[width=0.9\textwidth]{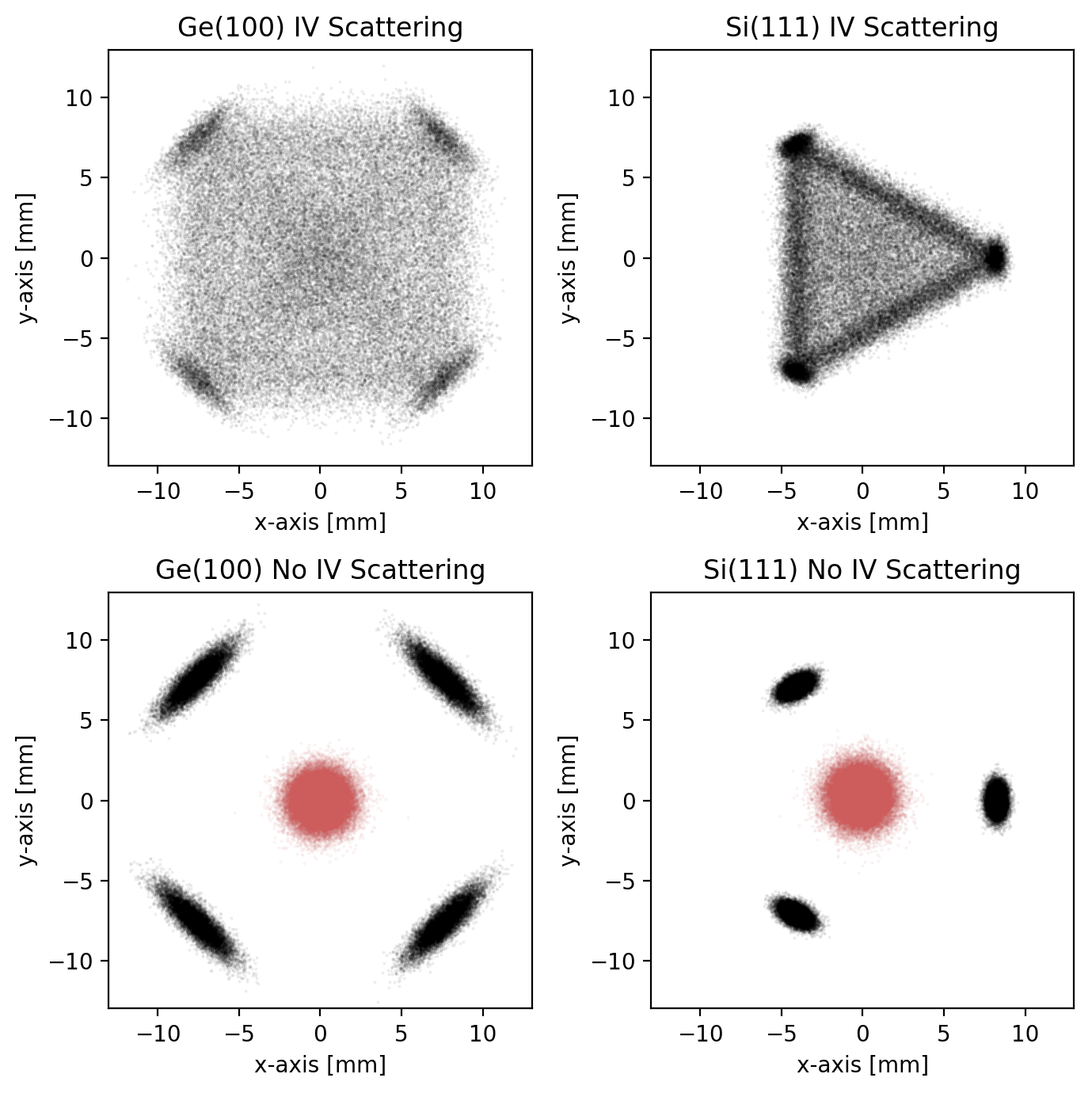}
	\caption{G4CMP simulations of electron transport (black) through 1.67\,cm thick Ge (left) and Si (right) crystals, both with (top) and without (bottom) intervalley (IV) scattering. In the former case (top), intervalley scattering tends to scatter electrons toward the center of the crystal and thus encourages transport to align with the applied field. For the cases without intervalley scattering (bottom), hole transport is also shown and appears as central clusters (red). The 0.5\,V/cm electric field is applied along the $z$ axis, which aligns with the $\langle$100$\rangle$ crystal axis in the case of Ge and the $\langle$111$\rangle$ axis for Si.}
	\label{fig:intervalley}
\end{figure*}

\begin{figure*}[t!]
    \centering
    \includegraphics[trim={0.225cm 0cm 0cm 0cm},clip,width=0.7\textwidth]{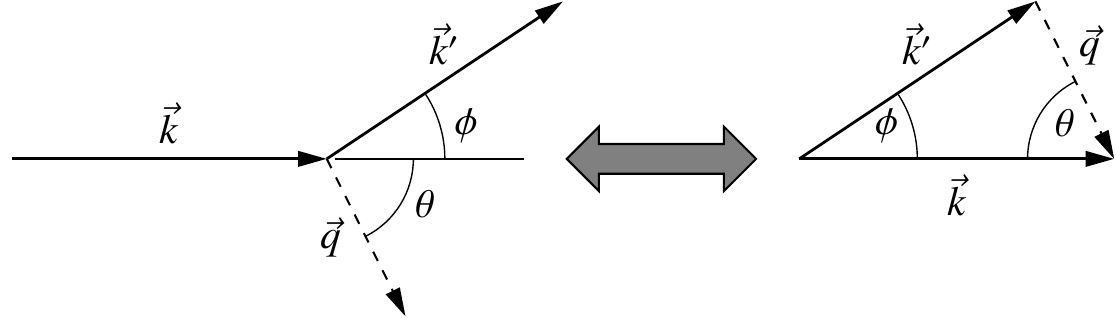}
    \caption{Elastic scattering of a charge carrier, with initial wavevector $\vec{k}$ and final wavevector $\vec{k}'$, off of the crystal lattice at an angle $\phi$, resulting in emission of an acoustic phonon with wavevector $\vec{q}$ at an angle $\theta$. Figure reproduced from Ref.~\cite{leman}.}
    \label{fig:scatter}
\end{figure*}

The combination of oblique propagation and intervalley scattering can produce distinct charge-collection patterns for electrons drifted across a Ge or Si crystal by an electric field.  
Again, optical phonons are not explicitly modeled as independent particles in G4CMP. Instead, G4CMP uses measured scattering rates to simulate charge transport. 
Intervalley scattering rates have been derived as a function of electric field for both Ge~\cite{moffatt,GeIV} (using the quadratic parameterization) and Si~\cite{SiIV1,SiIV2} (using the linear parameterization), and are used  
to set the scattering amplitudes in the G4CMP framework.  
The results of simulating electrons and holes propagating
through Ge and Si 
crystals are shown in Fig.~\ref{fig:intervalley}.  Simulations were performed both with (top row) and without (bottom row) the intervalley scattering process. A comparison between the two demonstrates which features arise from intervalley scattering versus oblique propagation. A uniform 0.5\,V/cm 
electric field was applied along the $z$ axis, aligned with the $\langle$100$\rangle$ crystal axis in Ge and the $\langle$111$\rangle$ axis in Si. As previously mentioned,   
electrons propagate along the L valleys in Ge which align with the $\langle$111$\rangle$ crystal axis (oblique to the applied field), producing four spatially distinct charge concentrations in the absence of intervalley scattering.  In Si, electrons propagate along the X valleys aligned with the $\langle$100$\rangle$ crystal axis, resulting in three spatially distinct charge concentrations. As shown in Fig.~\ref{fig:intervalley}, intervalley scattering tends to scatter electrons toward the center of the crystal and thus encourages transport to align with the applied field. 
The bottom panels also show hole transport simulated with the G4CMP framework. The broadening of the charge concentrations resulting from emission of NTL phonons is apparent. 
The dependence of these charge patterns on field strength is measured and simulated for Si in Ref.~\cite{SiIV2}.

\subsubsection{\label{sec:NTL}Neganov-Trofimov-Luke (NTL) phonon emission}
G4CMP can be used to simulate transport of charge carriers in a semiconductor crystal under application of an electric field. A numerical model for the electric field is provided as an input file, containing a triangulated mesh of electric-potential nodes specific to the device geometry and pre-calculated with a separate program (\textit{e.g.}, COMSOL~\cite{comsol}). The field causes charge carriers to drift and accelerate through the crystal, with their speed ultimately limited  
by emission of so-called NTL phonons~\cite{NTL1,NTL2,NTL3}, a charge-phonon scattering process analogous to Cherenkov radiation. One result of this phenomenon is that charge carriers -- both holes and electrons -- do not generally propagate along straight lines in semiconductors due to scattering with NTL phonons.
Because electrons and holes have different effective mass properties,  emission of NTL phonons is implemented differently for each type of charge carrier, as explained below. 

\paragraph{Holes}
In Ge and Si, the effective mass of a hole is well-approximated by an isotropic scalar. Holes therefore tend to propagate along the direction of the applied electric field.  
The anisotropic hole propagation discussed and measured in Refs.~\cite{SiIV1,SiIV2} is a relatively small effect that is not currently modeled in G4CMP.\footnote{The degeneracy of hole bands about the $\Gamma$ point makes a pseudo-linear transform like that done for electrons much more difficult, requiring a momentum-dependent effective-mass tensor. For more information see Ref.~\cite{kurinskyThesis}.}

Hole-phonon scattering is an elastic process, as shown in Fig.~\ref{fig:scatter}. From conservation of energy and momentum it follows that
\begin{equation}
\begin{split}
    k'^2 &= k^2 + q^2 - 2kq\cos{\theta},\\
    q &= 2(k\cos{\theta} - k_L),
\end{split}
\end{equation}
where $k_L = mv_L/\hbar$, $v_L$ is the magnitude of the longitudinal
phonon phase velocity, and $m$ is the effective mass of the hole \cite{Cabrera}.
With these constraints and the relationships from Fig.~\ref{fig:scatter}, one can solve for the angle $\phi$ of the scattered charge carrier:
\begin{equation}
    \cos{\phi} = \frac{k^2 - 2k_L(k\cos{\theta} - k_L) - 2(k\cos{\theta} -
k_L)^2}{k\sqrt{k^2 - 4k_L(k\cos{\theta} - k_L})}.
    \label{eq:scatterangle}
\end{equation}
Using Fermi's Golden Rule, the scattering rate is
\begin{equation}
    1/\tau = \frac{v_Lk}{3l_0k_L}\left(1-\frac{k_L}{k}\right)^3,
    \label{eq:rate}
\end{equation}
with an angular distribution
\begin{equation}
    P(k,\theta)\,d\theta =
    \frac{v_L}{l_0}\left(\frac{k}{k_L}\right)^2\left(\cos{\theta}-\frac{k_L}{k}\right)^2\sin{\theta}\,d\theta,
    \label{eq:ang-dist}
\end{equation}
where $0\le\theta\le\arccos{(k_L/k)}<\pi/2$, $l_0= \frac{\pi\hbar^4\rho}{2m^3C^2}$ is a characteristic scattering length, $\rho$ is the crystal density~\cite{leman}, and $C$ is the deformation potential constant~\cite{leman}.

\paragraph{Electrons}
Unlike the hole, the electron has a tensor effective mass in Ge and Si. As a result, some coordinate transformations need to be applied before propagating the electron through the crystal using the recipe outlined for hole propagation. For a coordinate system with one axis aligned with the principal axis of the conduction valley, the electron's equation of motion is
\begin{equation}
    \frac{eE_i}{m_i} = \frac{dv_i}{dt}\ ,
    \label{el_eq_mtn}
\end{equation}
where $i$ is the coordinate index, and $m_i$ is the mass along that coordinate axis (see below). 
To apply the same NTL-phonon emission recipe to electrons as holes, we apply a Herring-Vogt transformation\footnote{Following Ref.~\cite{leman}, the transformation is given in the basis where the $x$ component is aligned with the current valley axis and the other two directions ($y,\,z$) are perpendicular to the alignment direction and to each other.},
\begin{equation}
    T_{\rm HV} = \left( \begin{array}{ccc}
                    \sqrt{\frac{m_{\rm eff}}{m_x}} & 0 & 0 \\
                    0 & \sqrt{\frac{m_{\rm eff}}{m_y}} & 0 \\
                    0 & 0 & \sqrt{\frac{m_{\rm eff}}{m_z}}\end{array}\right),
 \label{HV}
\end{equation}
into a coordinate system in which the electron kinetic energy is independent of direction. In that space,
$v_i^* = v_i/\sqrt{m_{\rm eff}/m_i}$, where the effective mass $m_{\rm eff}$ is given by $3/m_{\rm eff} = 1/m_x +
1/m_y + 1/m_z$. For cubic crystals (including Si and Ge), the two masses perpendicular to the valley axis are degenerate, with the different masses conventionally denoted as $m_\parallel = m_x$ and $m_\perp = m_y = m_z$.  The specific mass values are substrate-dependent numerical constants (see Ref.~\cite{leman}). The electron's equation of motion in the rotated conduction valley frame is transformed to
\begin{equation}
    \frac{eE^*_i}{m_{\rm eff}} = \frac{dv_i^*}{dt}.
    \label{el_eq_mtn1}
\end{equation}
Following the application of the Herring-Vogt transformation, 
the same recipe that applies to holes for NTL-phonon emission can be followed for electrons~\cite{leman}.

\subsubsection{\label{sec:ct-ii}Charge trapping \& impact ionization}
As electrons and holes propagate through crystals, they may interact with 
various types of impurities, conventionally labelled $D$ for donor and $A$ for acceptor impurities.
At deeply cryogenic temperatures, the ``overcharged'' H$^-$-like impurity states ($D^-$ and $A^+$) 
are dominant~\cite{Gershenzon,Sundqvist}; only these impurity states are modeled in G4CMP. 
Several processes can occur depending on the sign of the charge and the type of impurity. Because we are concerned only about the propagating charges, and not the state of the impurity, G4CMP simplifies these into just two interaction modes: {\it charge trapping} and {\it impact ionization}~\cite{CT1,CT2}.  Both of these are described in detail in Ref.~\cite{Sundqvist}.

Charge trapping is a process in which a charge is captured by an impurity. Specifically, G4CMP models
\begin{eqnarray}
    \label{eq:etrap}
    \mathrm{e}^- D^0 &\rightarrow& D^-\ \mathrm{,}\\
    \label{eq:htrap}
    \mathrm{h}^+ A^0 &\rightarrow& A^+\ \mathrm{.}
\end{eqnarray}
The charge stops propagating and ceases producing NTL phonons via the process described in Sec.~\ref{sec:NTL}.  These traps are shallow compared to the bandgap, so the trapped charge is simply removed from the simulation with no released energy. 
For a device operated with an applied electric field, this trapping effect results in charge and phonon signals that are smaller than would otherwise be expected, with the reduction of the phonon signal depending on how far the charge propagated before it was trapped.

Impact ionization is a process in which the propagating charge ejects an additional charge from an impurity.
Specifically, G4CMP models
\begin{eqnarray}
    \label{eq:eII}
    \mathrm{e}^- D^- \rightarrow \mathrm{e}^- \mathrm{e}^- D^0 &{\rm or}& \mathrm{e}^- A^+ \rightarrow \mathrm{e}^- \mathrm{h}^+ A^0\ \mathrm{,}\\
    \label{eq:hII}
    \mathrm{h}^+ D^- \rightarrow \mathrm{h}^+ \mathrm{e}^- D^0 &{\rm or}& \mathrm{h}^+ A^+ \rightarrow \mathrm{h}^+ \mathrm{h}^+ A^0\ \mathrm{.}
\end{eqnarray}
For example, an electron may encounter a negative impurity and ionize it by ejecting an electron from it, 
or a hole may be created when an electron interacts with a positive impurity. 
The impurity becomes neutral, so no corresponding opposite charge is ejected, and the assumption of shallow traps means that the incident charge does not lose energy. The ejected charge begins propagating, and the original charge continues propagating; thus, both charges contribute to production of NTL phonons thereafter. This results in charge and phonon signals that are larger than would otherwise be expected, with the increase of the phonon signal depending on how far the initial charge propagated before it encountered the impurity.

In total, G4CMP uses six parameters to model this behavior, corresponding to two mean free paths for electron or hole trapping, respectively (Eqs.~\ref{eq:etrap} and \ref{eq:htrap}), and four mean free paths for either electrons or holes to produce an additional electron or hole via impact ionization (Eqs.~\ref{eq:eII} and \ref{eq:hII}).
These parameters generally vary from device to device, depending on crystal impurities and operating conditions, and must be specified by the user (see Table~\ref{tab:physics_config}). 

\begin{figure}[b!]
    \centering
    \includegraphics[trim={0.4cm 0.4cm 0.75cm 0.5cm},clip,width=0.48\textwidth]{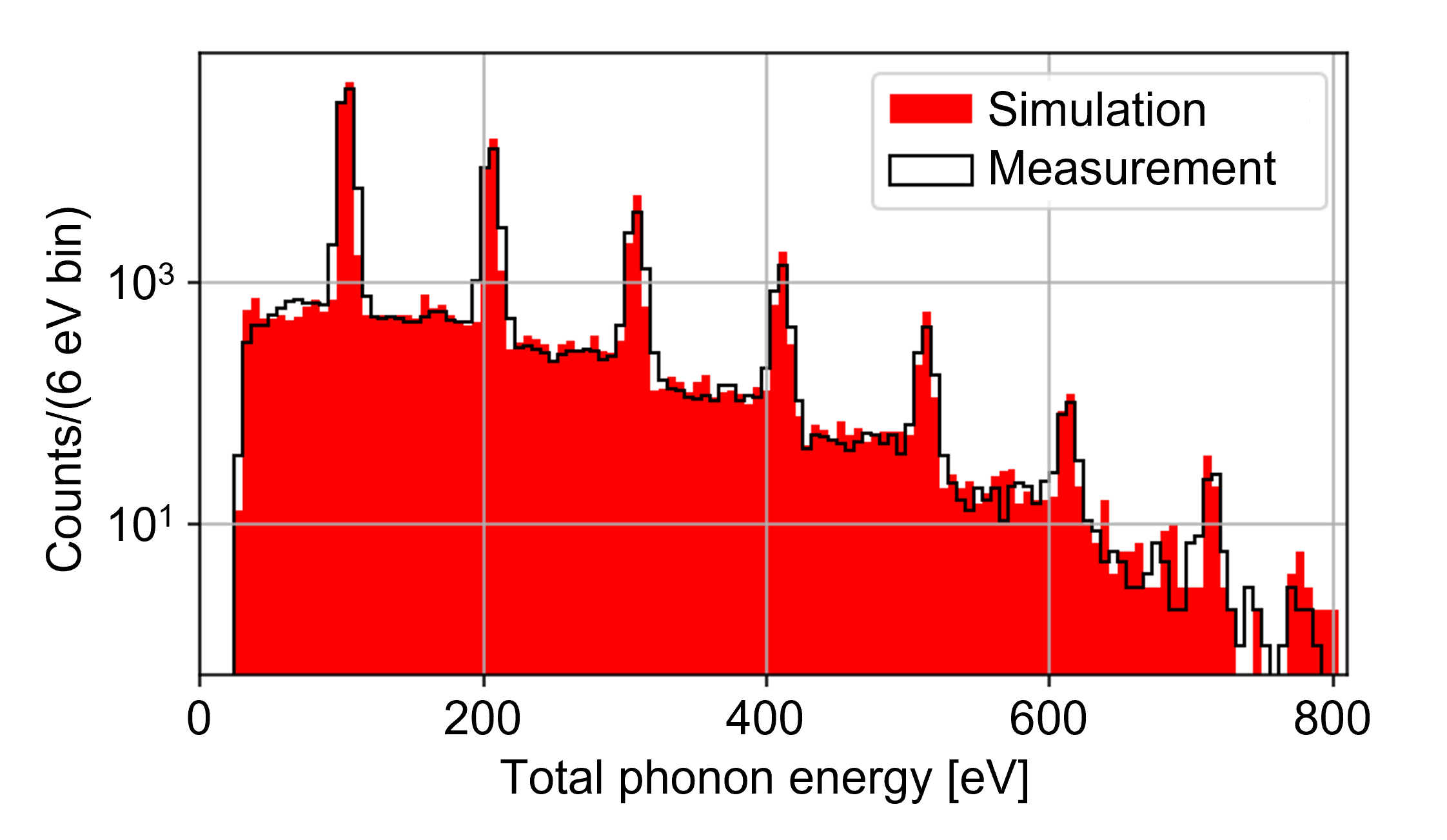}
    \caption{Simulated phonon-response energy spectrum (red) for a SuperCDMS HVeV detector operated with a 100\,V bias and exposed to 635\,nm (1.95\,eV) calibration photons, compared to the measured spectrum (black) from Ref.~\cite{SuperCDMS:2020ymb}. The peaks correspond to integer numbers of e$^-$/h$^+$ pairs. Events with energies between the peaks are attributed to experiencing the effects of charge trapping and impact ionization. The parameters used to simulate these processes in G4CMP were obtained by fitting the model in Ref.~\cite{CT2} to the measured spectrum.}
    \label{fig:ctii}
\end{figure}

As an example, we show a calibration spectrum measured with a SuperCDMS HVeV device~\cite{SuperCDMS:2020ymb} in Fig.~\ref{fig:ctii}, compared to a simulation of the spectrum with G4CMP. The peaks in the spectrum correspond to the total phonon energies of integer numbers of e$^-$/h$^+$ pairs for a 100\,V detector bias, from a single pair at $\sim$100\,eV to seven pairs at $\sim$700\,eV. 
Charge trapping and impact ionization can result in phonon energies corresponding to nonintegral numbers of e$^-$/h$^+$ pairs, tending to fill in the regions between the peaks. Because the phonon resolution of this device is more than sufficient to cleanly resolve gaps between the e$^-$/h$^+$ pair peaks, the spectral shape is highly sensitive to and can be used to fit for this device's trapping and impact-ionization parameters~\cite{CT1,CT2}.  
As Fig.~\ref{fig:ctii} demonstrates, using the best-fit parameters in G4CMP  
allows for a reasonable simulation of this device's measured response.

\subsubsection{Charge carrier drift speed}
\begin{figure}[t!]
    \centering
    \includegraphics[trim={0.4cm 0.4cm 0.4cm 0.4cm},clip,width=0.48\textwidth]{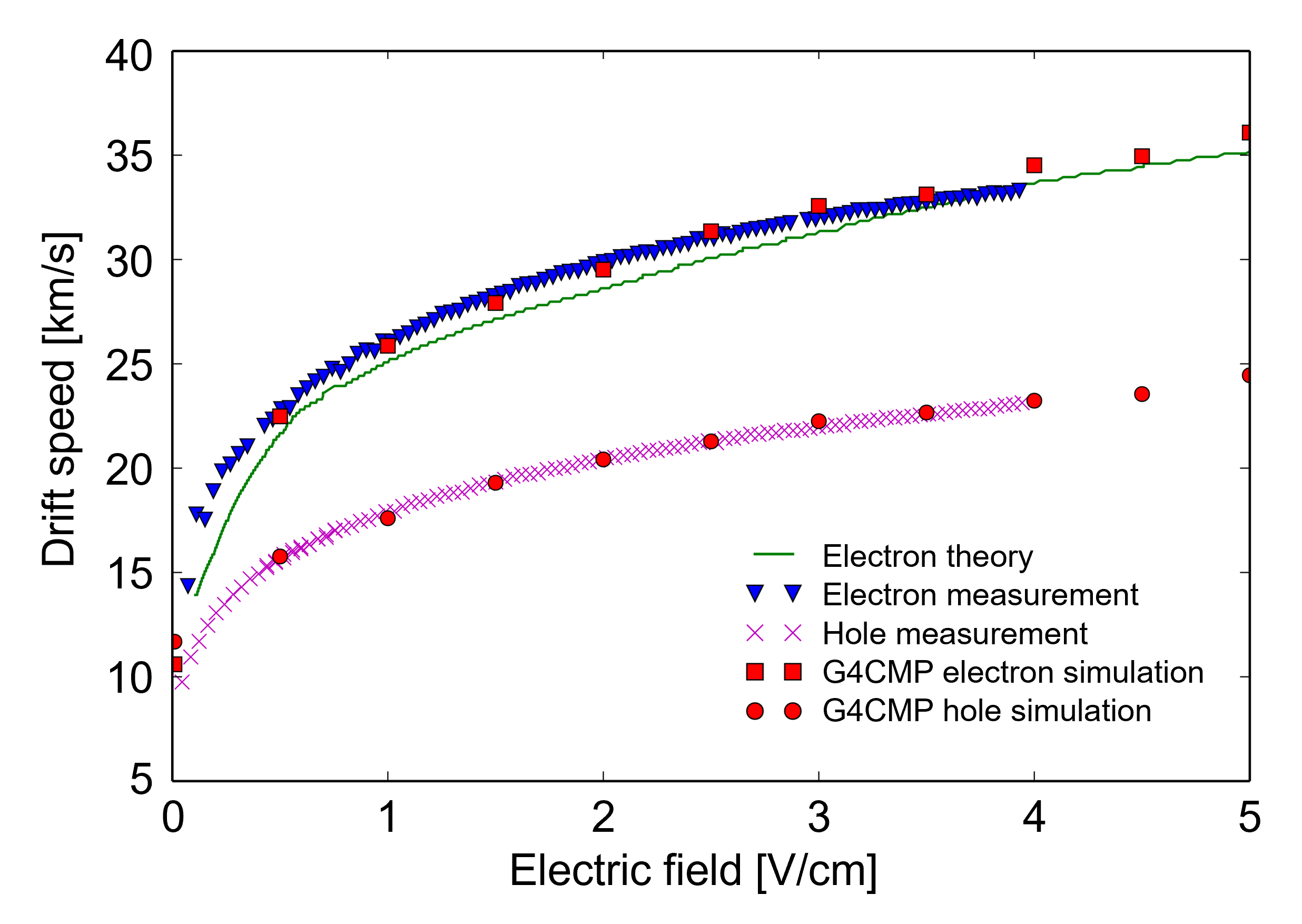}
    \caption{G4CMP-simulated drift speeds in Ge versus applied electric-field strength for electrons (red squares) and holes (red circles), compared to experimental data from Ref.~\cite{Sundqvist} for electrons (blue triangles) and holes (magenta $\times$'s)  and to the theoretical model from Ref.~\cite{Fortuna} (green curve). Figure adapted from Ref.~\cite{arXiv:1403.4984}.}
    \label{fig:results}
\end{figure}

An important observable for charge transport under an applied electric field is drift speed. Comparison of simulated and measured drift speeds is thus a good test that the G4CMP charge-transport processes are working as intended.  
For different values of the electric field, a maximum drift
speed of the charge carrier is reached, at which point energy from the electric field goes into emission of NTL phonons. Figure \ref{fig:results} shows the average speeds of electrons and holes drifted through a Ge crystal with $\mathcal{O}$(V/cm) electric fields. The intervalley scattering rates for the G4CMP simulations are estimated using  the theoretical model in Ref.~\cite{Fortuna}. The simulated results incorporate the effects of all the charge transport processes described in this section, yielding good agreement with experimental data \cite{Sundqvist}. For Si, G4CMP's intervalley scattering rates are based on the measurements in Refs.~\cite{SiIV1,SiIV2}, including the option to select either of the two modeling methods reported in those papers.

\subsection{\label{sec:Quasiparticles}Superconducting quasiparticle processes}
Geometry surfaces may be configured in G4CMP to include modeling of phonon absorption by superconducting films and associated sensor response (see Sec.~\ref{sec:surfaces}). 
To model quasiparticles in superconducting films, 
we have implemented a ``lumped'' version~\cite{leman} of Kaplan's model for phonon-quasiparticle behavior~\cite{Kaplan} in a class named {\tt G4CMPKaplanQP}. 
Our model traces the energy flow from a substrate phonon incident on the film to the amount of energy recorded by a sensor, including quasiparticle production, quasiparticle decay, and phonons re-emitted into the substrate, without tracking individual particles directly in the film.   
As implemented, energy exchange between phonons and quasiparticles is evaluated iteratively and the process effectively occurs instantaneously -- any phonons re-emitted into the substrate are assigned the same tracking timestamp as the incident phonon. 
Consequently, energy delocalization via quasiparticle diffusion is not yet captured by this package.
Future extensions of G4CMP will include modeling of quasiparticle recombination and concomitant phonon emission in order to improve the generality of the toolkit beyond the originally intended dense arrays of athermal phonon sensors.

Included in the model is the possibility for the initial incident phonon to be reflected back into the substrate without absorption (\textit{i.e.}, no quasiparticles are produced), with a probability~\cite{leman} 
\begin{equation}
    P_{\rm escape}(E_{ph}) =
    \exp\left(-\frac{2\cdot 2d}{\lambda(E_{ph})}\right)\ ,
    \label{eq:escapepdf}
\end{equation}
where $E_{ph}$ is the phonon energy, and $d$ is the film thickness.\footnote{Note that G4CMP includes another mechanism for phonons to reflect at the substrate-film interface, which is covered in Sec.~\ref{sec:surfaces}.}
One factor of 2 in the exponent accounts for the phonon needing to traverse (at least) twice the thickness of the film to reflect back into the substrate.  The other factor of 2 results from integrating over the possible angles of incidence of the initial phonon with the film.
Here, $\lambda(E_{ph})$ is the mean free path for phonons as a function of energy and phonon lifetime $\tau$:
\begin{equation}
    \lambda(E_{ph}) = 
    \frac{v_{\rm sound}\tau}
      {1 + \delta\tau\cdot(E_{ph}/\Delta - 2)}\ ,
    \label{eq:phononmfp}
\end{equation}
where $v_{\rm sound}$ is the speed of sound in the film material, $\Delta$ is the superconducting bandgap of the film material, and $\delta\tau$ is a dimensionless slope of the phonon lifetime {\it vs.\@} energy.

Phonons with energy $E_{ph} < 2\Delta$ may be collected directly by the sensor with a user-defined probability {\tt subgapAbsorption} (see Sec.~\ref{sec:surfaces}).
If such a phonon is not collected, it is reflected back into the substrate.

If the phonon is neither reflected nor is directly absorbed, its energy is transferred into a pair of Bogoliubov quasiparticles, breaking a Cooper pair in the superconducting film.
The energy is shared between the two quasiparticles as $E_{qp}$ and $E_{ph}-E_{qp}$, where the energy is chosen using an accept-reject loop according to the function~\cite{leman,Agnese:2017mfx}
\begin{equation}
    f(E_{qp}|E_{ph}) = 
      \frac{E_{qp}\left(E_{ph}-E_{qp}\right)+\Delta^2}
        {\sqrt{\left(E_{qp}^2-\Delta^2\right)
        \left(\left(E_{ph}-E_{qp}\right)^2-\Delta^2\right)}}\ .
    \label{eq:qpenergypdf}
\end{equation}
Quasiparticles with energies below $3\Delta$ are collected onto the sensor and removed from further consideration.
Higher-energy quasiparticles will scatter in the film, effectively decaying by losing energy to a phonon in the process: $E'_{qp} = E_{qp} - E_{ph}$, where $E_{ph}$ is chosen using an accept-reject loop according to the function~\cite{leman,Agnese:2017mfx}
\begin{equation}
    f(E_{ph}|E_{qp}) =
    \frac{E_{ph}\,(E_{qp}-E_{ph})^2\,(E_{ph}- \frac{\Delta^2}{E_{qp}})}
      {\sqrt{E_{ph}^2 - \Delta^2}}\ .
    \label{eq:phenergypdf}
\end{equation}
The initial quasiparticle energy is replaced with the new value $E'_{qp}$, and the new phonon energy is stored. 
Phonons created by quasiparticle scattering are passed through the same procedure described above, with possibilities for emission into the substrate, direct collection on the sensor, or conversion into additional quasiparticles. 

This procedure is iterated in {\tt G4CMPKaplanQP} until there are no remaining quasiparticles; the total energy collected on the sensor, together with the remaining phonon energies, are returned.
The phonons from quasiparticle scattering which are re-emitted into the substrate are assigned random polarizations chosen according to the steady-state phonon populations~\cite{Tamura4} (see Sec.~\ref{sec:PhononProcesses}) and random directions. A momentum-preserving implementation of this down-conversion process is described in Ref.~\cite{lehmann}, which can be used to produce initial phonon and quasiparticle distributions of known momenta for spatially accurate transport simulations in superconductors.\footnote{A python implementation of this simulation procedure can be found at~\url{https://github.com/benvlehmann/scdc}.} Work is ongoing to implement these processes in future upgrades to G4CMP.
  

\section{\label{sec:features}G4CMP features and functionality}
The interface between G4CMP and {\sc Geant4} is in the form of a {\it physics builder} for adding the G4CMP processes to an existing user simulation.  The package also provides a standalone {\it physics list} for running a solid-state-only simulation.  Relevant properties for a given crystal volume are specified via material data files included in the G4CMP package. There are a set of global parameters, listed in \ref{sec:config}, which can be used to customize the performance of G4CMP; some of these are mentioned specifically below. Interactions between charge carriers or phonons and the surface of a crystal volume, including interfaces with sensor volumes, are handled with {\sc Geant4}-style surface properties.

\subsection{Physics configuration}\label{sec:physicsconfig}
As discussed in Sec.~\ref{sec:Production}, G4CMP accesses the energy deposited by radiation in {\sc Geant4} to create e$^-$/h$^+$ pairs and prompt phonons.
Configuring this behavior involves multiple parameters and data files in G4CMP, which can be set using {\sc Geant4} macro commands.
Crystal properties, phonon velocities, charge-carrier effective masses, scattering rates, {\it etc.}, are defined for each material (see Sec.\ \ref{sec:materials}).
G4CMP currently includes predefined properties for Si and Ge, under the CrystalMaps directory, referenced via the environment variable \texttt{\$G4LATTICEDIR}.  This directory may be redirected by changing the environment variable or via the macro command \texttt{/g4cmp/LatticeData}.

For energy deposits involving heavy ions (sometimes called ``nuclear recoils''), either as atoms within the crystal or as projectiles (\textit{i.e.}, \texttt{G4Track} instances), the Lindhard model~\cite{Lindhard} is used to compute the partitioning of energy for the production of e$^-$/h$^+$ pairs versus the creation of prompt phonons. Two different models are provided in the G4CMP library: Lindhard \& Robinson~\cite{Robinson} ({\tt lindhard})
and Lewin \& Smith~\cite{LewinSmith} ({\tt lewin}). 
  User applications can specify the model using \texttt{/g4cmp/NIELPartition} with one of these name strings.

The Fano factor for e$^-$/h$^+$ pair production from an energy deposit is a property of the crystal material and is set with the \texttt{fanoFactor} parameter in \texttt{config.txt}. 
This feature enables fluctuation in the number of charge carriers produced ($N_{\mathrm{eh}}$) for a given energy deposit $E_{\rm dep}$. An exact number of pairs
$N_{\mathrm{eh}} = E_{\rm dep}/E_{\rm pair}$ 
can instead be generated by setting \texttt{/g4cmp/enableFanoStatistics false}, where $E_{\rm pair}$, the average energy required to create an e$^-$/h$^+$ pair, is a material-dependent G4CMP parameter.

\subsection{\label{sec:surfaces}Defining surface properties}
How phonons and charge carriers behave at the surface of a solid-state volume is configured analogously to optical surfaces in {\sc Geant4}~\cite{Geant-B},
including reflection, absorption or termination, and transmission.
G4CMP provides classes to define a general boundary for a logical volume ({\tt G4CMP\-Logi\-cal\-Skin\-Sur\-face}) or the boundary between two physical ({\it placement}) volumes ({\tt G4CMP\-Logi\-cal\-Bor\-der\-Sur\-face}).
These classes are specific to G4CMP and may be instantiated for volumes without conflicting with {\sc Geant4} optical surfaces.
As with optical surfaces, each G4CMP surface instance must be supplied with a {\tt G4\-Sur\-face\-Pro\-per\-ty} instance, either as a constructor argument or via {\tt SetSurfaceProperty()}.  
For convenience, G4CMP provides a {\tt G4CMPSurfaceProperty} subclass which includes a constructor that takes all the necessary arguments for specifying reflection or termination of both phonons and charge carriers, as well as frequency-dependent reflection or down-conversion of phonons, and which will fill material property tables with those arguments.

Because the devices being modeled often include a sensor or ``electrode'' for collecting signals, {\tt G4CMP\-Sur\-face\-Pro\-per\-ty} supports registration of user-defined electrode classes as subclasses of {\tt G4CMPV\-Elec\-trode\-Pat\-tern};
one electrode instance for phonons and one for charge carriers may be registered to each surface.\footnote{For such an application, the geometry should include small volumes representing each sensor attached to the substrate, and electrodes should be registered to the border surfaces between those sensor volumes and the substrate.}
When an electrode is registered, the {\tt absProb} material property is used to determine whether the electrode collects the phonon or charge carrier or passes it to the surface reflection code.
The material properties table for the surface is passed (by pointer) into the electrode class automatically.  User applications may define their own properties to be used by their electrode subclass and add those properties to the surface's associated table.

\begin{figure*}[t!]
    \centering
    \includegraphics[width=0.45\textwidth]{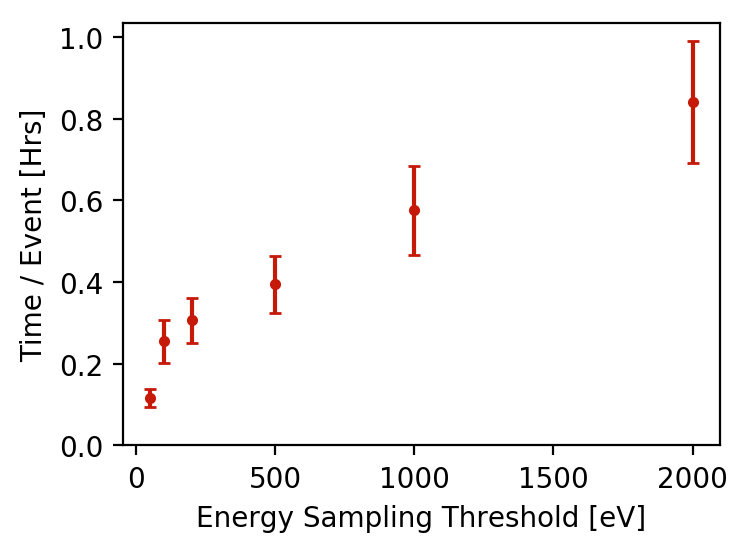}\qquad
    \includegraphics[width=0.45\textwidth]{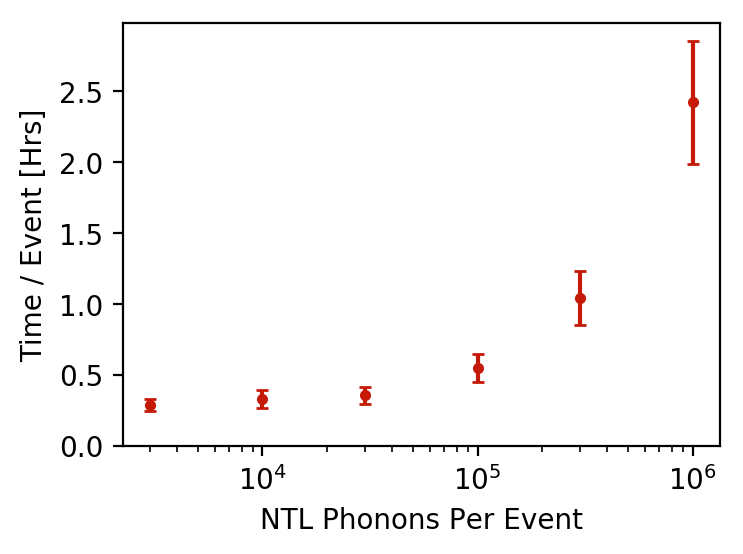}
    \caption{Processing time per event in hours for a 10~keV energy deposit from a gamma-ray interaction in a Ge crystal versus down-sampling parameters. \textbf{Left:} Dependence on the energy sampling threshold, with the max number of NTL phonons per event set to  $10^5$. \textbf{Right:} Dependence on the max number of NTL phonons per event, with the energy sampling threshold set to 1000~eV.}
    \label{fig:CPU}
\end{figure*}

The G4CMP library provides {\tt G4CMPPhononElectrode} as a concrete implementation of a phonon detection electrode.
This electrode is intended to be attached to the border surface between the crystal and a volume representing a superconducting film (such as aluminum) with some energy-absorbing and response functionality such as a transition-edge sensor (TES).  The behavior of such a surface device is modeled with the {\tt G4CMPKaplanQP} class, called from the electrode.
{\tt G4CMPPhononElectrode} expects a set of properties that specify the response of the film, including 
its thickness ({\tt filmThickness}), superconducting energy gap $\Delta$ ({\tt energyGap}\footnote{Note that {\tt energyGap} is defined as the superconducting bandgap $\Delta$, {\em not} the energy $2\Delta$ needed to break a Cooper pair.}),
phonon lifetime $\tau$ ({\tt phononLifetime}), dimensionless slope $\delta\tau$ of the phonon lifetime {\it vs.\@} energy ({\tt PhononLifetimeSlope}; see Eq.\ \ref{eq:phononmfp}),
speed of sound ({\tt vSound}), 
and minimum energy for the code to have a quasiparticle radiate a phonon ({\tt lowQPLimit}, which multiplies $\Delta$). 
An optional property is available to represent the additional behavior of the energy absorber in the electrode: 
the probability ({\tt subgapAbsorption}) for an incident phonon to be directly absorbed, which, for example, might correspond to the effective area of the absorber relative to the superconducting film.

\subsection{\label{sec:materials}Adding new materials}
G4CMP comes with options for tracking phonons and charge carriers in either silicon or germanium substrates. 
Implementation of a new material requires creation of a configuration file to be added as \texttt{CrystalMaps/$<$name$>$/config.txt},
either in the G4CMP repository or a user-specific directory pointed to with the \texttt{\$G4LATTICEDIR} environment variable. 
Most of the parameters currently included in the existing \texttt{config.txt} files are needed, so the Si or Ge file can be used as a template for creating a file for a new material.
See also \ref{sec:matconfig} for an explanation of each parameter.

\subsection{Computation: resources and down-sampling}
The potentially wide range of energies in G4CMP simulations -- from MeV radiation interactions to sub-meV phonons -- can result in extremely large  
numbers of tracks and steps per track. 
A typical event starting with a few-keV gamma ray may take a CPU day or more to simulate. 
To make simulations tractable, G4CMP provides parameters for {\it down-sampling} the production of e$^-$/h$^+$ pairs and prompt phonons from energy deposits, and for the number of NTL phonons to be tracked in each event. Figure~\ref{fig:CPU} shows some examples of the CPU time associated with differing choices of down-sampling parameters.

Down-sampling is specified using a sampling threshold $E_{\rm sampling}$ ({\tt /g4cmp/samplingEnergy}), or via fixed fractions for charges ({\tt /g4cmp/produceCharges}) and prompt phonons ({\tt /g4cmp/producePhonons}). 
The former is used with each energy deposit to compute a sampling fraction $E_{\rm sampling}/E_{\rm deposit}$ to use if $E_{\rm deposit} > E_{\rm sampling}$. 
In either case, a weight $W = 1/{\rm fraction}$ is assigned to each created track (and its subsequent interactions).

For NTL phonons, a ``soft maximum'' number per event $N^{\rm max}_{\rm NTL}$ can be specified with the macro command {\tt /g4cmp/maxLukePhonons}.
The number of e$^-$/h$^+$ pairs and the detector bias are used to estimate how many NTL phonons may be generated, assuming an average of 2\,meV per phonon. 
This estimate is combined with $N^{\rm max}_{\rm NTL}$ to set a fraction and weight for NTL emission.
A default value of $N^{\rm max}_{\rm NTL} = 10000$ is set in G4CMP.

The computational resources needed for G4CMP physics simulations are nontrivial.  As presented in Fig.~\ref{fig:CPU}, processing times are on the order of an hour per event, and simulation jobs often entail processing of tens of thousands to many millions of events. 
Thus, users should consider investigation of down-sampling and validation of simulation results as part of the application development process.


\section{\label{sec:examples}Applications of G4CMP}
There are several reports of the application of G4CMP toward understanding charge and phonon transport in chip-based, superconducting devices. These reports extend the application of G4CMP beyond the original intent of modeling the response of solid-state cryogenic dark-matter detectors~\cite{Agnese:2017mfx} and the charge-transport results in Refs.~\cite{SiIV1,SiIV2} (see Sec.~\ref{sec:InterValley}). An initial example explored the utility of modeling phonon transport in a silicon chip instrumented with kinetic-inductance detectors (KIDs)~\cite{Martinez:2018ezx}. This study evaluated interfacial phonon-energy-absorption efficiencies in comparison to response data collected from two KID-containing devices. A second modeling application evaluated the use of normal-metal pads or chip-trenching to mitigate transport of phonons (produced by cosmic rays) into TESs and KIDs~\cite{minami2020irradiation}. This work focused on the planned LiteBIRD satellite mission that would be located at Earth's L2 Lagrange point to study B-mode polarization of the cosmic microwave background radiation. More recently, G4CMP was employed to understand charge diffusion in devices hosting charge-sensitive transmon qubits~\cite{Wilen:2020lgg}. This report helped reveal the \emph{correlated} nature of environmentally-induced errors (\textit{e.g.}, due to gamma and cosmic-ray interactions) in arrays of qubits sharing a common substrate.

The reader is encouraged to review these applications in Refs.~\cite{Martinez:2018ezx,minami2020irradiation,Wilen:2020lgg} which provide examples of the utility of G4CMP for enhancing the understanding of chip-based superconducting devices. In the remainder of this section, we expand the list of example applications, further demonstrating the breadth of cases and questions that G4CMP can address. These examples are kept brief and focused on their specific application of G4CMP. More detailed device and scientific results related to these examples appear in the literature and may appear in future reports.

\subsection{\label{sec:hvev}Chip-scale cryogenic dark matter sensor}
SuperCDMS HVeV detectors~\cite{Romani:2017iwi,Ren:2020gaq}, comprising superconducting sensors on chip-scale silicon substrates, are sensitive to $\mathcal{O}$(eV) energy deposits and are used to search for dark matter with low-energy thresholds~\cite{SuperCDMS:2018mne,SuperCDMS:2020ymb}. 
These devices are fabricated with a dense pattern of athermal phonon sensors -- quasiparticle-trap-assisted electrothermal-feedback transition-edge sensors (QETs)~\cite{QET, QET2} -- with high efficiency for phonon energy collection.
 
As mentioned in Sec.~\ref{sec:ct-ii}, a key characteristic is the response to few-integer multiples of e$^-$/h$^+$ pairs when such a device is operated with a bias voltage of $\mathcal{O}$(100\,V), resulting in production of NTL phonons which amplify the phonon signal  (see Fig.~\ref{fig:ctii}). Underlying the HVeV data analysis is a detailed understanding of the phonon pulse response. A 635\,nm (1.95\,eV) laser produces events of known origin, which are used to create a pulse-shape template for reconstructing the phonon energy of events of unknown origin.

The SuperCDMS collaboration has developed a Detector Monte Carlo (DMC) software framework to model the detector response to energy deposits, including initial interactions modeled by {\sc Geant4}, charge and phonon simulation modeled by G4CMP, and an additional software package to model the digitized sensor readout.
G4CMP provides the underlying physics processes needed to simulate production and tracking of charges and phonons, including their interactions  at the surface of the Si substrate (see Sec.\ \ref{sec:processes}).
The phonon-electrode configuration described in Sec.\ \ref{sec:surfaces} is used to model the collection of phonon energy by the QETs.
Finally, SuperCDMS has developed a software package 
which models the electrothermal response of the QETs and readout electronics to phonon energy, and which produces digitized waveforms equivalent to the experimental readout. 
This code is implemented in the form of coupled differential equations for the network of QETs, equivalent to the TES simulation in Ref.~\cite{leman}.

\begin{figure}[t!]
    \centering
    \includegraphics[trim={0.2cm .2cm 0.2cm 0.2cm},clip,width=0.48\textwidth]{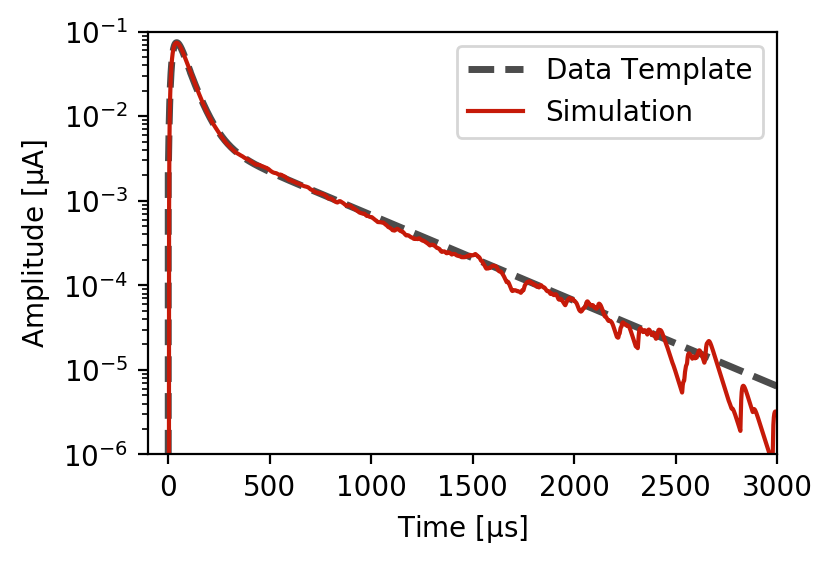}
    \caption{Parameter-optimized phonon pulse response  simulated for an HVeV device using G4CMP and the SuperCDMS Detector Monte Carlo (red solid), compared to the pulse template  constructed from HVeV laser calibration data (black dashed).}
    \label{fig:HVeVfit}
\end{figure}

Initial comparisons of data from a laser-pulsed HVeV device and the output from the DMC model showed that the simulated pulses were almost twice as large in amplitude as the measured pulses. The pulse rise times and peak positions also were not well-matched to data. 
More telling, the presence of a second exponential fall time in the measured pulses was not reproduced by the simulation model. These differences are attributed to incorrect parameters and deficiencies in the DMC model. Three DMC parameters were found to significantly influence the characteristics of the simulated pulse response:
\begin{itemize}
    \item The QET absorption probability determines if a phonon is absorbed into a QET aluminum fin, independent of the Si substrate's specular \textit{vs.}\ diffuse surface reflectivity;
    \item The superconducting gap ($\Delta$) of the aluminum determines if an absorbed phonon has sufficient energy ($2 \Delta$) to break a Cooper pair (if not, the phonon is reflected diffusely back into the substrate); and
    \item The sub-gap absorption parameter sets the probability for low-energy phonons to be directly absorbed by the QET's tungsten TES.
\end{itemize}

\begin{figure*}[t!]
	\centering
    \includegraphics[width=0.99\textwidth]{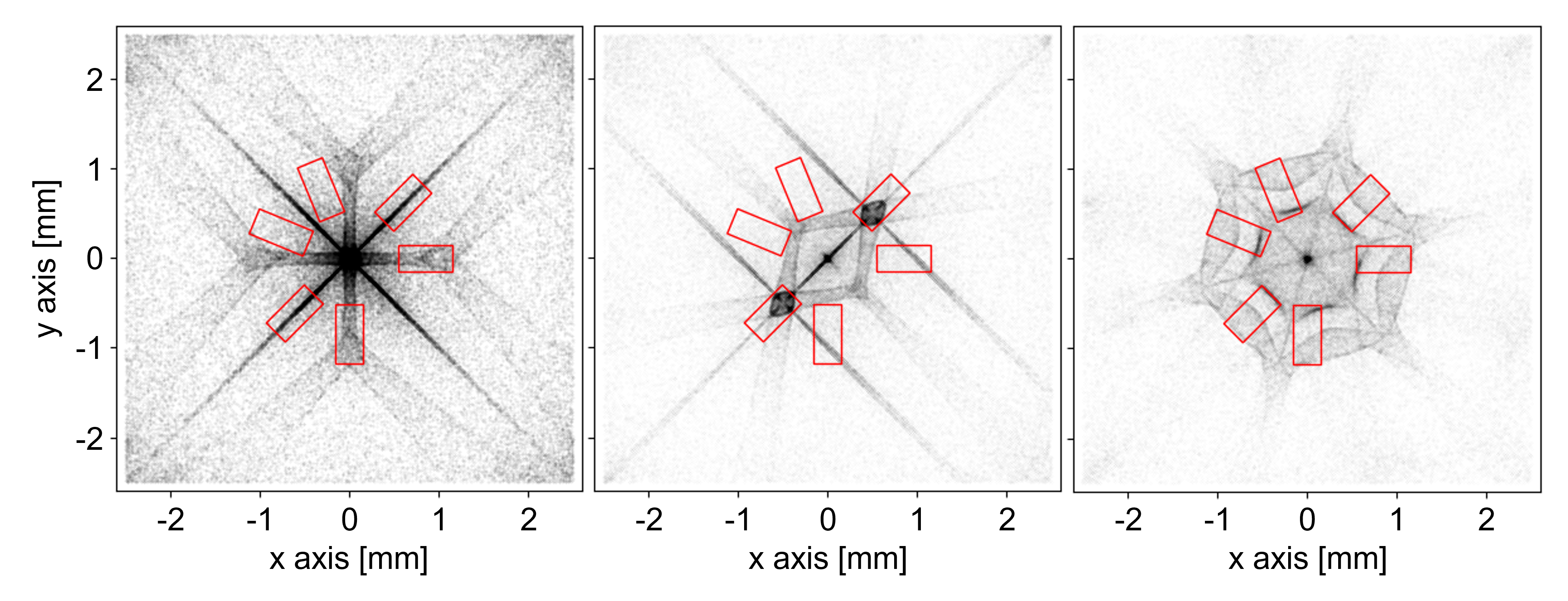}
	\caption{Patterns of phonon caustics simulated using G4CMP for a $5 \times 5$\,mm$^2$,  350\,$\mu$m thick silicon chip cut at different crystal orientations: (from left to right) $\langle$100$\rangle$, $\langle$110$\rangle$, $\langle$111$\rangle$ orientations.
    The red rectangles indicate proposed positions of sensors to collect phonon energy.}
	\label{fig:Sichipped}
\end{figure*}

Figure~\ref{fig:HVeVfit} compares the pulse template constructed from HVeV laser calibration data to the simulated pulse response after DMC parameter tuning.  
Specifically, this result was generated by using an optimal filter fitting procedure, which accounts for uncertainties due to noise and returns a $\chi^2$ value that measures goodness-of-fit between a simulated pulse and the data template. Through iterative assessment of this $\chi^2$ metric as a function of parameter values, a final set of DMC parameters was found that achieves a $\chi^2$ minimum. The QET absorption and sub-gap absorption parameters arrived at values that are physically motivated, whereas the favored aluminum superconducting energy gap is unphysically large. This unphysical value implies that there is a deficiency in the model likely related to the time and energy dependence of quasiparticle propagation in the QETs.

\subsection{\label{sec:qischip}Superconducting device for caustics measurement}
Direct detection of phonon caustics patterns can potentially be used to validate the phonon transport models implemented in G4CMP, which could help to optimize the spatial configuration of on-chip sensors. 
These caustics could be probed in chip-scale devices comprised of spatially-localized superconducting sensors fabricated on a single-crystal substrate, such as Si or Ge.
Established pair-breaking sensor technologies could be utilized, such as microwave kinetic inductance detectors (MKIDs)~\cite{day2003} or quantum capacitance detectors (QCDs)~\cite{Echternach2018}, provided that a localized source of phonons with sufficient emission at energies $>2\Delta$ can be integrated on-chip.
To demonstrate this, sensors could be placed radially at varying angles relative to the crystal axes, with the origin defined by the position of the phonon source.
Pulsed injection of phonons, followed by high-bandwidth detection of the sensor response, would resolve anisotropic phonon transport and the subsequent generation of quasiparticles via phonon absorption in the sensors. 
The G4CMP package was employed in the development and layout of devices having these goals.

Figure \ref{fig:Sichipped} shows a G4CMP simulation of phonon transport for a feasible device geometry, with a localized phonon source co-located on the same surface as the sensors. A $5\times 5$\,mm$^2$, 350\,$\mu$m thick Si crystal is used, with a polished bare surface on the bottom (for specular reflection of the phonons) and a fully metalized top surface with 100\% absorption. Phonons of 0.9\,meV are injected from the center of the top surface isotropically downward (at 90$^\circ$ half angle) into the Si crystal. The red rectangles show proposed sensor locations overlaid on top of the simulated caustics patterns which result from phonon transport in the Si crystal.

\subsection{\label{sec:qpdensity}Modeling device trenching and phonon transport}

One factor impacting superconducting device performance is an undesirable excess of quasiparticles in the superconducting films. Various environmental factors 
can deposit energy into the substrate (see, \textit{e.g.}, Refs.~\cite{Vepsalainen:2020trd,Wilen:2020lgg,Cardani:2020vvp,McEwen:2021wdg,stress}),  
which in turn generates phonons that can break Cooper pairs in the superconductor. These pair-breaking phonons were shown as a source of decoherence in superconducting quantum circuits and devices \cite{PatelQP}. The G4CMP toolkit provides a platform to simulate various methods to mitigate substrate phonon transport into superconducting circuits, such as a ``phonon sink'' in the form of a normal-metal film anchored to the substrate  
or etching away the substrate material to create a pseudo-island that effectively isolates sensitive superconducting elements from the surrounding substrate \cite{Iaia2022,minami2020irradiation}. Simulation of these mitigation techniques is a potentially valuable design tool for optimizing device performance when energy transfer from substrate phonons is a concern. 

\begin{figure}[t!]
    \centering
    \includegraphics[width=0.48\textwidth]{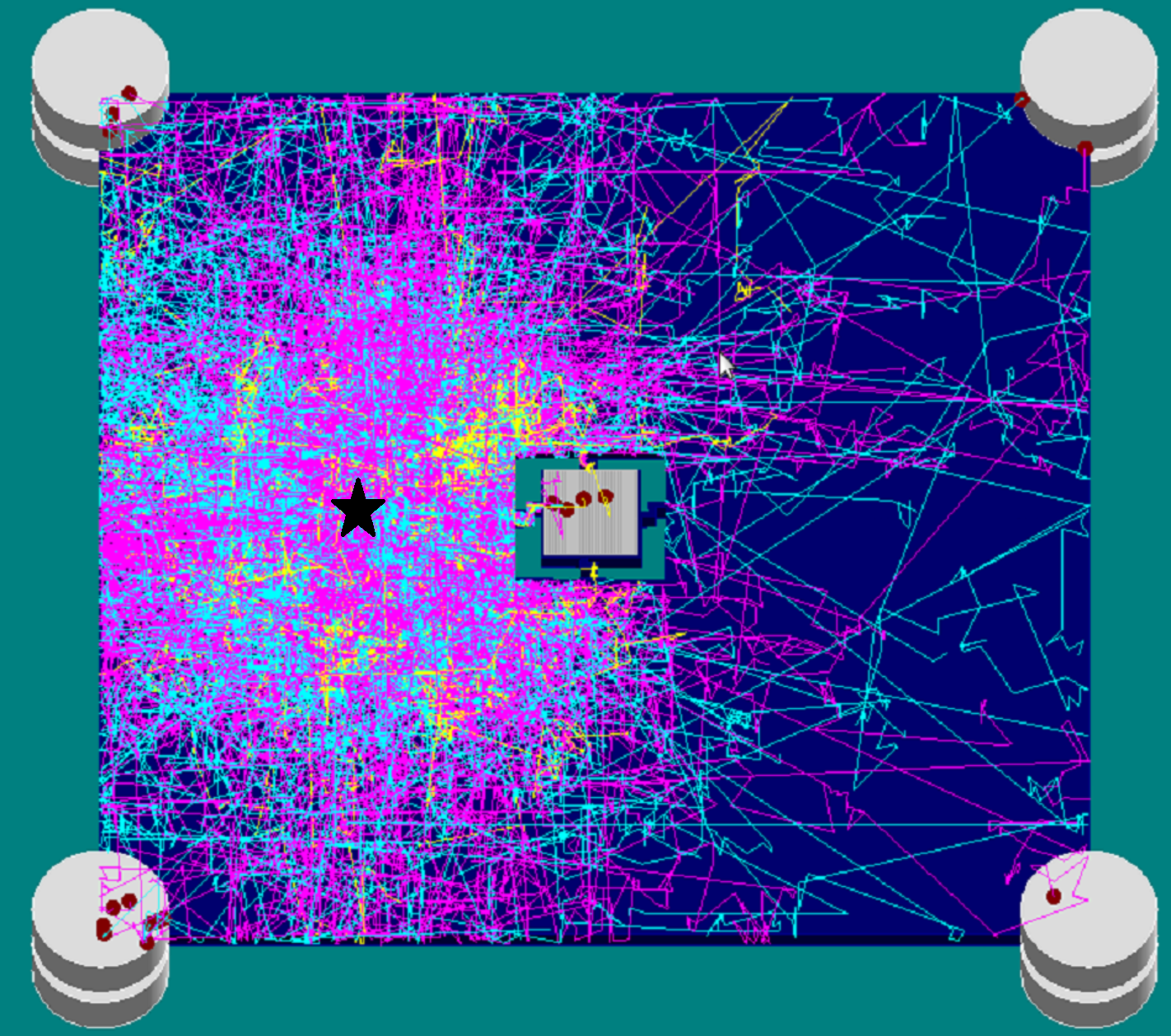}
    \caption{Simulation model of a superconducting sensor 
    with substrate etching to limit phonon propagation. The sensor (aluminum film on central island) is $2\times2$\,mm$^{2}$ and is connected to the surrounding Si substrate via small ``legs.'' Phonon tracks corresponding to the substrate's acoustic modes -- fast transverse, slow transverse, and longitudinal -- are shown in cyan, magenta, and yellow, respectively. Dark red circles indicate phonon absorption sites. Phonons were generated by throwing photons at a location on the substrate surface outside the central island (black star).}
    \label{fig:G4CMPEtchGeometry}
\end{figure}

We present here an example simulation that focuses on etching away (or ``micro-machining'') the substrate near a sensor with the goal of limiting phonon propagation from the substrate into the superconducting film. Based on the ``P1'' MKID device in Ref.~\cite{Martinez:2018ezx}, the simulation geometry is shown in Fig.~\ref{fig:G4CMPEtchGeometry} and includes a superconducting sensor (aluminum film) centered on top of a silicon substrate, legs that connect the central island to the surrounding substrate, and device clamps at the four substrate corners. Phonons are generated as a result of photons thrown at the outer substrate. The phonon tracks are shown for the corresponding, color-coded acoustic modes: cyan for fast transverse, magenta for slow transverse, and yellow for longitudinal.

The shape, thickness, and position of the legs were varied for these simulations. Figure~\ref{fig:G4CMPEtchGeometry} shows the ``dog-leg'' geometry. A simpler, straight-leg version was also simulated for comparison. For each design, a parametric sweep was performed for each leg geometry with varying leg thicknesses. The presence of metal films on the top and bottom surfaces of the legs  
was also simulated to explore phonon absorption. 
The simulation results indicate that the dog-leg geometry with metal films added to the legs was the most effective way to isolate the island-based sensor from phonons. Varying the thickness of the legs between $1$\,$\mu$m and $1$\,mm shows that phonon hits at the central sensor decrease with decreasing leg thickness (for each leg type), as expected. This latter result is true whether or not phonon sinking metal was included on the surfaces of the legs.


\section{\label{sec:summary}Summary and outlook}
In summary, the {\sc Geant4} Condensed Matter Physics (G4CMP) software package is used by and developed for simulation of sub-Kelvin cryogenic devices employing superconducting sensors or circuits. Originally developed to model the response of solid-state dark-matter detectors, G4CMP is now being used for a broader range of applications, including modeling of superconducting sensors for satellite missions and superconducting qubits for quantum computing. 
In this article, we provide descriptions and demonstrations of the underlying physics processes included for simulation of phonon and charge transport in semiconductor substrates, as well as details for the phonon-sensitive superconducting electrode that is currently implemented as part of the package.  
To highlight the potential breadth of application, we present  
examples ranging from {\sc Geant4}-incorporated device-response models (\textit{i.e.}, HVeV device in Sec.~\ref{sec:hvev}) to pure G4CMP phonon transport with native {\sc Geant4} output (\textit{i.e.}, caustics device in Sec.~\ref{sec:qischip} and trenched substrate in Sec.~\ref{sec:qpdensity}).

Research within the many user communities is driving development of new features for the G4CMP package. 
We are actively developing several features that may appear in future releases.  To support studies of quasiparticle populations in the superconducting circuitry of quantum devices, we are developing particle transport and interaction models for  superconducting films. These new processes for temporal and spatial tracking of quasiparticles and phonons in superconducting films are analogous to the processes already included in G4CMP for transport of phonons and charges in semiconductor substrates. 
Additionally, we are adding more realistic features to the {\tt G4CMPKaplanQP} sensor response model described in Sec.~\ref{sec:Quasiparticles}, informed by the development work on quasiparticle-phonon interactions just mentioned, and by measurements and modeling of sensor response (\textit{e.g.}, as in Sec.~\ref{sec:hvev}).
To support simulation of a wider range of devices, we are also working with user groups to develop material properties tables (see Sec.~\ref{sec:materials} and \ref{sec:matconfig}) for additional substrates (beyond silicon and germanium) and superconducting films (besides aluminum).

As G4CMP provides the underlying physics processes, it is both possible and encouraged for users to develop specific geometries, add new materials, and implement new electrode models as needed to simulate their devices.
Users may contribute to the project through the GitHub repository~\cite{G4CMP_code_repo} by creating issues, or by forking and submitting pull requests.  Users may also request to be added to the project directly as GitHub ``Contributors'' and may then submit code modifications via feature branches directly to the repository. 

\section{\label{sec:acknowledgements}Acknowledgements}
The original development of G4CMP took place more than a decade ago~\cite{arXiv:1403.4984}. Many have contributed to the package since that time. Our intention with this article was to bring all contributors together with several ``user groups'' for an updated presentation of the code and breadth of potential application. Thus, this work was truly a multi-collaboration endeavor. The authors thank each of their respective sponsoring agencies for making possible the development and use of the G4CMP software package.

We gratefully acknowledge support from the U.S.\ Department of Energy (DOE) Office of High Energy Physics (HEP), the DOE Office of Nuclear Physics (NP), the National Science Foundation (NSF), the Natural Sciences and Engineering Research Council of Canada (NSERC), the Canada First Research Excellence Fund through the Arthur B.\ McDonald Canadian Astroparticle Physics Research Institute, and the Mitchell Institute for Fundamental Physics and Astronomy at Texas A\&M University. 
This work was supported in part under grants from the DOE HEP QuantISED and DOE NP Quantum Horizons programs. This research was enabled in part by support provided by SciNet (\href{https://www.scinethpc.ca/}{scinethpc.ca}) and the Digital Research Alliance of Canada (\href{https://alliancecan.ca/en}{alliancecan.ca}), and portions were conducted with the advanced computing resources provided by Texas A\&M High Performance Research Computing.   
This research was supported by an appointment to the Intelligence Community Postdoctoral Research Fellowship Program at Massachusetts Institute of Technology administered by Oak Ridge Institute for Science and Education (ORISE) through an interagency agreement between the U.S.\ DOE and the Office of the Director of National Intelligence (ODNI).  
Y.-Y.\,Chang was supported in part by a Taiwanese Ministry of Education Fellowship and by a Caltech J.\ Yang Fellowship. 
Funding was received by the Deutsche Forschungsgemeinschaft (DFG, German Research Foundation) under the Emmy Noether Grant No.\ 420484612. 
Fermilab is operated by the Fermi Research Alliance, LLC, under Contract No.\ DE-AC02-37407CH11359 with the
U.S.\ DOE. Pacific Northwest National Laboratory is a multi-program national laboratory operated for the U.S.\ DOE by Battelle Memorial Institute under Contract No.\ DE-AC05-76RL01830. SLAC is operated under Contract
No.\ DE-AC02-76SF00515 with the U.S.\ DOE. Work at MIT Lincoln Laboratory is supported under Air Force Contract No.\ FA8702-15-D-0001. Any opinions, findings, conclusions or recommendations expressed in this material are those of the authors and do not necessarily reflect the views of the U.S.\ Government.


\appendix
\section{\label{sec:config}Global configuration parameters}
To configure a G4CMP simulation, a number of parameters are available ranging from the mechanics of controlling the simulation job-processing, debugging, CPU efficiency, and down-sampling to specifying the geometry configuration and physics processes. These configuration parameters are succinctly listed in Tables~\ref{tab:job_config}--\ref{tab:physics_config}.

\begin{table*}
  \centering
  \caption{Job control parameters.\label{tab:job_config}}
 \begin{tabular}{>{\tt}p{0.3\textwidth}l}
  \hline
  \hline
  \rm Environment variable \\[-0.2em]
  \rm \quad Macro command & Value/action \\
  \hline
  G4LATTICEDATA [D] & Directory with lattice configs \\[-0.2em]
  \quad /g4cmp/LatticeData [D] \\[0.5em]
  G4CMP\_DEBUG [L] & Enable diagnostic messages \\[-0.2em]
  \quad /g4cmp/verbose [L] $>$0 \\[0.5em]
  G4CMP\_HIT\_FILE [F] & Write e$^-$/h$^+$ hit locations to ``F'' \\[-0.2em]
  \quad /g4cmp/HitsFile [F] \\
  \hline
  \hline
\end{tabular}
\end{table*}
  
\begin{table*}
  \centering
  \caption{Debugging and CPU efficiency parameters.\label{tab:debug_config}}
  \begin{tabular}{>{\tt}p{0.35\textwidth}l}
  \hline
  \hline
  \rm Environment variable \\[-0.2em]
  \rm \quad Macro command & Value/action \\
  \hline
  G4CMP\_EH\_BOUNCES [N] & Maximum e$^-$/h$^+$ reflections \\[-0.2em]
  \quad /g4cmp/chargeBounces [N] \\[0.5em]
  G4CMP\_PHON\_BOUNCES [N] & Maximum phonon reflections \\[-0.2em]
  \quad /g4cmp/phononBounces [N] \\[0.5em]
  G4CMP\_EMIN\_PHONONS [E] & Minimum energy to track phonons \\[-0.2em]
  \quad /g4cmp/minEPhonons [E] eV \\[0.5em]
  G4CMP\_EMIN\_CHARGES [E] & Minimum energy to track charges \\[-0.2em]
  \quad /g4cmp/minECharges [E] eV \\[0.5em]
  G4CMP\_MIN\_STEP [S] & Force minimum step S\*L0 \\[-0.2em]
  \quad /g4cmp/minimumStep [S] $>$0 \\
  \hline
  \hline
\end{tabular}
\end{table*}

\begin{table*}
  \centering
  \caption{Down-sampling (reweighting) parameters.\label{tab:downsampling_config}}
  \begin{tabular}{>{\tt}p{0.4\textwidth}l}
  \hline
  \hline
  \rm Environment variable \\[-0.2em]
  \rm \quad Macro command & Value/action \\
  \hline
  G4CMP\_MAKE\_PHONONS [R] & Fraction of phonons from energy deposit \\[-0.2em]
  \quad /g4cmp/producePhonons [R] \\[0.5em]
  G4CMP\_MAKE\_CHARGES [R] & Fraction of charge pairs from energy deposit \\[-0.2em]
  \quad /g4cmp/produceCharges [R] \\[0.5em]
  G4CMP\_LUKE\_SAMPLE [R] & Fraction of generated Luke phonons \\[-0.2em]
  \quad /g4cmp/sampleLuke [R] \\[0.5em]
  G4CMP\_MAX\_LUKE [N] & Soft maximum Luke phonons per event \\[-0.2em]
  \quad /g4cmp/maxLukePhonons [N] \\[0.5em]
  G4CMP\_SAMPLE\_ENERGY [E] & Energy above which to down-sample \\[-0.2em]
  \quad /g4cmp/samplingEnergy [E] eV \\[0.5em]
  G4CMP\_COMBINE\_STEPLEN [L] & Combine hits below step length \\[-0.2em]
  \multicolumn{2}{l}{\tt \quad /g4cmp/combiningStepLength [L] mm} \\
  \hline
  \hline
\end{tabular}
\end{table*}

\begin{table*}
  \centering
  \caption{Geometry configuration parameters.\label{tab:geometry_config}}
  \begin{tabular}{>{\tt}p{0.375\textwidth}l}
    \hline
    \hline
  \rm Environment variable \\[-0.2em]
  \rm \quad Macro command & Value/action \\
  \hline
  G4CMP\_CLEARANCE [L] & Minimum distance of tracks from boundaries \\[-0.2em]
  \quad /g4cmp/clearance [L] mm \\[0.5em]
  G4CMP\_VOLTAGE [V] & Apply uniform +Z voltage \\[-0.2em]
  \quad /g4cmp/voltage [V] volt !=0: \\[0.5em]
  G4CMP\_EPOT\_FILE [F] & Read mesh field file ``F'' \\[-0.2em]
  \quad /g4cmp/EPotFile [F] V=0 \\[0.5em]
  G4CMP\_EPOT\_SCALE [F] & Scale the potentials in EPotFile by factor F\\[-0.2em]
  \quad /g4cmp/scaleEPot [F] V=0 \\[0.5em]
  G4CMP\_MILLER\_H & Miller indices for lattice orientation \\[-0.2em]
  G4CMP\_MILLER\_K & \\[-0.2em]
  G4CMP\_MILLER\_L & \\[-0.2em]
  \quad /g4cmp/orientation [h] [k] [l] \\[0.5em]
  G4CMP\_TEMPERATURE & Device/substrate/{\it etc.} temperature \\[-0.2em]
  \quad /g4cmp/temperature [T] K \\
      \hline
    \hline
\end{tabular}
\end{table*}

\begin{table*}
  \centering
  \caption{Physics process configuration.\label{tab:physics_config}} 
  \begin{tabular}{>{\tt}p{0.5\textwidth}l}
  \hline
  \hline
  \rm Environment variable \\[-0.2em]
  \rm \quad Macro command & Value/action \\
  \hline
  G4CMP\_USE\_KVSOLVER &  Use eigensolver for K-Vg mapping \\[-0.2em]
  \quad /g4mcp/useKVsolver [t,f] \\[0.5em]
  G4CMP\_FANO\_ENABLED & Apply Fano statistics to input ionization \\[-0.2em]
  \multicolumn{2}{l}{\tt \quad /g4cmp/enableFanoStatistics [t,f]} \\[0.5em]
  G4CMP\_NIEL\_FUNCTION & Select NIEL partitioning function \\[-0.2em]
  \multicolumn{2}{l}{\tt \quad /g4cmp/NIELPartition [LewinSmith, Lindhard]} \\[0.5em]
  G4CMP\_CHARGE\_CLOUD & Create charges in sphere around location \\[-0.2em]
  \quad /g4cmp/createChargeCloud [t,f] \\[0.5em]
  G4CMP\_IV\_RATE\_MODEL & Select intervalley rate parametrization \\[-0.2em]
  \multicolumn{2}{l}{\tt \quad /g4cmp/IVRateModel [IVRate, Linear, Quadratic]} \\[0.5em]
  G4CMP\_ETRAPPING\_MFP & Mean free path for electron trapping \\[-0.2em]
  \quad /g4cmp/eTrappingMFP [L] mm \\[0.5em]
  G4CMP\_HTRAPPING\_MFP & Mean free path for charge hole trapping \\[-0.2em]
  \quad /g4cmp/hTrappingMFP [L] mm \\[0.5em]
  G4CMP\_EDTRAPION\_MFP & MFP for electron-trap ionization by e$^-$ \\[-0.2em]
  \multicolumn{2}{l}{\tt \quad /g4cmp/eDTrapIonizationMFP [L] mm} \\[0.5em]
  G4CMP\_EATRAPION\_MFP & MFP for hole-trap ionization by e$^-$ \\[-0.2em]
  \multicolumn{2}{l}{\tt \quad /g4cmp/eATrapIonizationMFP [L] mm} \\[0.5em]
  G4CMP\_HDTRAPION\_MFP & MFP for electron-trap ionization by h$^+$ \\[-0.2em]
  \multicolumn{2}{l}{\tt \quad /g4cmp/hDTrapIonizationMFP [L] mm} \\[0.5em]
  G4CMP\_HATRAPION\_MFP & MFP for hole-trap ionization by h$^+$ \\[-0.2em]
  \multicolumn{2}{l}{\tt \quad /g4cmp/hATrapIonizationMFP [L] mm} \\
  \hline
  \hline
\end{tabular}
\end{table*}

\section{\label{sec:matconfig}Lattice definition parameters}
To define crystalline materials that can support charge and phonon transport within the G4CMP framework, the user must enter values for the parameters in Tables~\ref{tab:lattice_config}--\ref{tab:intervalley_model_config} into a file named \texttt{config.txt} for the desired material. See Sec.~\ref{sec:materials} for additional details.
The G4CMP package includes pre-defined data files  for silicon and germanium (\texttt{CrystalMaps/*/config.txt}).  Users can find the specific parameter values in these files.

\begin{table*}
   \centering
   \caption{Lattice configuration parameters used in {\tt config.txt}.\label{tab:lattice_config}}
\begin{tabular}{>{\tt}llll}
\hline \hline
  {\rm Keyword} & Arguments & Value type(s)             & Units \\ \hline
  amorphous & --none--  & Polycrystalline solid     & \\
  cubic   & a         & Lattice constant          & length \\
  tetragonal & a c    & Lattice constants         & length \\
  hexagonal  & a c    & Lattice constants         & length \\
  orthorhombic & a b c & Lattice constants        & length \\
  rhombohedral & a $\alpha$ & Lattice const., angle  & length, deg/rad \\
  monoclinic & a b c $\alpha$ & Lattice const., angle & length, deg/rad \\
  triclinic & a b c $\alpha$ $\beta$ $\gamma$ & Lattice const., angle & length, deg/rad \\
  stiffness & i j val & Indices 1--6, elasticity   & pressure (Pa, GPa) \\
  Cij       & i j val & Indices 1--6, elasticity   & Pa, GPa \\
  \hline \hline
\end{tabular}
\end{table*}

\begin{table*}
   \centering
   \caption{Phonon parameters used in {\tt config.txt}.\label{tab:phonon_config}}
\begin{tabular}{>{\tt}llll}
\hline \hline
  {\rm Keyword} & Arguments & Value type(s)             & Units \\ \hline
  beta    & val       & Scattering parameters     & Pa, GPa \\
  gamma   & val       & (see Ref.~\cite{Tamura3}) & Pa, GPa \\
  lambda  & val       &                           & Pa, GPa \\
  mu      & val       &                           & Pa, GPa \\
  dyn     & $\beta$ $\gamma$ $\lambda$ $\mu$ & All four parameters & Pa, GPa \\
  scat    & B         & Isotopic scattering rate  & second$^3$ (s3) \\
  decay   & A         & Anharmonic decay rate     & second$^4$ (s4) \\
  decayTT & frac      & Fraction of L$\rightarrow$TT decays  & \\
  LDOS    & frac      & Longitudinal density of states & sum to unity \\
  STDOS   & frac      & Slow-transverse density of states & \\
  FTDOS   & frac      & Fast-transverse density of states & \\
  Debye   & val       & Debye energy for phonon primaries & E, T, Hz \\
  vsound  & Vlong     & Sound speed (longitudinal) & m/s \\
  vtrans  & Vtrans    & Sound speed (transverse)   & m/s \\
  \hline \hline
\end{tabular}
\end{table*}

\begin{table*}
   \centering
   \caption{Charge carrier parameters used in {\tt config.txt}.\label{tab:charge_config}}
\begin{tabular}{>{\tt}llll}
\hline \hline
  {\rm Keyword} & Arguments & Value type(s)             & Units \\ \hline
  bandgap & val       & Bandgap energy             & energy (eV) \\
  pairEnergy & val    & Energy taken by e-h pair   & energy (eV) \\
  fanoFactor & val    & Spread of e-h pair energy  & \\
  l0\_e   & len       & Electron scattering length & length \\
  l0\_h	  & len       & Hole scattering length     & length \\
  hmass   & m$_h$       & Effective mass of hole   & m/m$_e$ \\
  emass   & m$_{xx}$ m$_{yy}$ m$_{zz}$ & Electron mass tensor & (same) \\
  valley  & $\theta$ $\phi$ $\psi$\ & Euler angles     & angle (deg/rad) \\
  \hline \hline
\end{tabular}
\end{table*}

\clearpage

\begin{table*}
   \centering
   \caption{Intervalley scattering matrix element parameters used in {\tt config.txt}.\label{tab:intervalley_config}}
\begin{tabular}{>{\tt}llll}
\hline \hline
  {\rm Keyword} & Arguments & Value type(s)             & Units \\ \hline
  epsilon & e/e0      & Relative permittivity     & \\
  neutDens & N        & Number density of neutron impurities & /volume \\
  alpha    & val      & Non-parabolicity of valleys & energy$^{-1}$ (/eV) \\
  acDeform & val      & Acoustic deformation potential & energy (eV) \\
  ivDeform & val val ... & Optical deformation potentials & eV/cm \\
  ivEnergy & val val ... & Optical phonon thresholds     & energy (eV) \\
  \hline \hline
\end{tabular}
\end{table*}

\begin{table*}
   \centering
   \caption{Intervalley scattering model parameters used in {\tt config.txt}.\label{tab:intervalley_model_config}}
\begin{tabular}{>{\tt}llll}
\hline \hline
  {\rm Keyword} & Arguments & Value type(s)             & Units \\ \hline
  ivModel     & name & IVRate (matrix), Linear or Quadratic   & string \\
  ivLinRate0  & val & Constant term in linear IV expression ($b$ in Eq.~\ref{eq:ivlinear})   & Hz \\
  ivLinRate1  & val & Linear term in linear IV expression ($m$ in Eq.~\ref{eq:ivlinear})    & Hz \\
  ivLinPower  & exp & Exponent ($\alpha$ in Eq.~\ref{eq:ivlinear}): rate = Rate0 + Rate1$\cdot$E$^{\rm exp}$ & none \\
  ivQuadRate  & val & Coefficient for quadratic IV expression ($A$ in Eq.~\ref{eq:ivquad}) & Hz \\
  ivQuadField & val & Minimum field for quadratic IV expression ($E_0$ in Eq.~\ref{eq:ivquad}) & V/m \\
  ivQuadPower & exp & Exponent ($\alpha$ in Eq.~\ref{eq:ivquad}): rate = Rate$\cdot$(E$^2$+Field$^2$)$^{({\rm exp}/2)}$ & none \\
  \hline \hline
\end{tabular}
\end{table*}

\clearpage
\bibliography{bibliography}

\end{document}